\documentclass[prb,onecolumn,superscriptaddress]{revtex4}
\usepackage{graphicx}
\usepackage{bm}
\usepackage{amsmath}

\newcommand{\kk}{{\bf k}}

\begin{document} \title{Input-output theory of cavities in the ultra-strong coupling regime:
\\
the case of a time-independent vacuum Rabi frequency}

\author{Cristiano Ciuti}
\email{ciuti@lpa.ens.fr} \affiliation{Laboratoire Pierre Aigrain,
Ecole Normale Sup\'erieure, 24, rue Lhomond, 75005 Paris, France}

\author{Iacopo Carusotto}
\affiliation{BEC-CNR-INFM and Dipartimento di Fisica, Universit\`a
di Trento, I-38050 Povo, Italy}

\begin{abstract}

We present a full quantum theory for the dissipative dynamics of an
optical cavity in the ultra-strong light-matter coupling regime, in
which the vacuum Rabi frequency is a significant fraction of the
active electronic transition frequency and the anti-resonant terms of the
light-matter coupling play an important role.
In particular, our model can be applied to the case of intersubband
transitions in doped semiconductor quantum wells embedded in a
microcavity.
The coupling of the intracavity photonic mode and of the electronic
polarization to the external, frequency-dependent, dissipation baths
is taken into account by means of quantum Langevin equations in the
input-output formalism.
In the case of a time-independent vacuum Rabi frequency, exact analytical
expressions for the operators are obtained, which
allows us to characterize the quantum dissipative response of the
cavity to an arbitrary initial condition (vacuum, coherent
field, thermal excitation).
For a vacuum input in both the photonic and electronic polarization
modes, the ground state of the cavity system is a two-mode squeezed
vacuum state with a finite population in both photonic and electronic
modes. These excitations are however virtual and can
not escape from the cavity: for a vacuum input, a vacuum output is
found, without any trace of the intracavity squeezing.
For a coherent photonic input the linear optical response spectra
(reflectivity, absorption, transmission) have been studied, and
signatures of the ultra-strong coupling have been identified in
the asymmetric and peculiar anticrossing of the polaritonic eigenmodes.
Finally, we have calculated the electroluminescence spectra in
  the case of an incoherent electronic input: the emission
    intensity in the ultra-strong coupling regime results
    significantly enhanced as compared to the case of an isolated
    quantum well without a surrounding cavity.
\end{abstract}

\date{\today} \maketitle

In recent years, the study of cavity quantum electrodynamics\cite{CQED,Haroche} has
been a subject of an intense and fruitful research. In particular,
a considerable deal of activity has been devoted to the so-called
strong light-matter coupling regime. This regime is achieved when
the vacuum Rabi frequency $\Omega_R$ of an electronic excitation
(i.e., the Rabi frequency associated to electric field vacuum
fluctuations) exceeds the frequency broadening of the photonic and
electronic excitations.
In the case of atoms in high-finesse cavities (either in the
optical~\cite{CQED} or in the microwave~\cite{Haroche} domains), the
strong coupling occurs thanks to the strong spatial localization of the
electromagnetic field and the very high quality factors (as
high as $10^8$) of the cavity mode and the atomic resonance, but the vacuum
Rabi frequency $\Omega_R$ still remains a tiny fraction of the active
transition frequency $\omega_{eg}$.
%(at most $\Omega_R/\omega_{eg}
%\approx 10^{-??}$ CHECK LITERATURE NOT ONLY FOR SINGLE ATOM CQED...).
In the case of semiconductor planar microcavities
\cite{Weisbuch_PRL,Dini_PRL}, the quality factors are by far less
impressive (typically not much larger than $Q \approx 10^3$), but the
strong coupling can be still comfortably obtained thanks to the strong
photon confinement and the large electric dipole moment of the active
electronic transitions in the quantum well. Both these facts
cooperate to give a much larger vacuum Rabi frequency. Using
inter-subband transitions in doped semiconductor quantum
wells\cite{Dini_PRL,Liu,Dupont,Aji_APL,SST},
it is even possible  to achieve an
unprecedented ultra-strong light-matter coupling
regime\cite{Ciuti_vacuum}, in which
the vacuum Rabi frequency $\Omega_R$ becomes comparable to the
intersubband electronic transition $\omega_{12}$.
In this regime, the standard rotating wave approximation is no longer
valid, and anti-resonant terms of the light-matter
interaction start playing a significant role.
In particular, it has been shown in Ref.\onlinecite{Ciuti_vacuum} that
the ground state
of an isolated cavity is a two-mode squeezed vacuum containing
 a finite number of virtual and correlated photons and
  intersubband excitations.

In the present paper, a full quantum description of the system is
developed, which now explicitely includes the coupling to dissipation
baths.
In the specific case of intersubband excitations in planar semiconductor
microcavities considered here,
the dissipation baths are not only responsible for damping rates which
are typically as large as 5-10 \% of the transition
frequency~\cite{Dini_PRL,Aji_APL,SST}, but  also provide the way of
exciting and observing the cavity dynamics.
The photonic mode is coupled to external world mostly because of the
finite reflectivity of the cavity mirrors, while the intersubband
transition is coupled to other excitations in the semiconductor
material, e.g. acoustic and optical phonons, and free carriers in
levels other than the ones involved in the considered transition.
In particular, the coupling to this electronic bath allows one
to electrically excite the intersubband transitions.

The theory here developed is based on the so-called input-output
formalism~\cite{Collett,Reynaud,Walls}, in which the dynamics of the
electronic polarization and cavity photonic fields is described
in terms of quantum Langevin equations for the two coupled quantum fields.
Differently from previous treatments, we  have to take into
account here all the anti-resonant terms of the vacuum Rabi coupling, which forces us for
consistency to keep track in an exact way of the frequency dependence
of the dissipation baths.

The paper is organized as follows.
In Sec. \ref{introduction}, we introduce the model Hamiltonian for the
considered system and
for the baths. In Sec. \ref{Langevin}, the quantum Langevin equations
for the intracavity photonic and electronic polarization fields are
derived. Their exact solution for the intracavity operators as well as
for the output operators is given in Sec.\ref{solutions}. The case
of a vacuum input both for the photonic and polarization fields is
analyzed in Sec. \ref{ground}, where the properties of the ground
state of the system are characterized and the presence of
squeezing in the  intra-cavity fields is pointed out.
The question of the observability or not of this squeezing in
the output field is addressed in Sec. \ref{visibility}, where we
predict that no squeezing can be observed in the output modes unless
the input is itself squeezed.

After the brief discussion of single- and double-sided cavities
  of Sec.\ref{double_section}, the main features of the linear optical
spectra (reflectivity, absorption, transmission) under optical
excitation are considered in Sec.\ref{reflection}. The
electroluminescence emission spectra under an electronic
excitation are studied in Sec.\ref{electroluminescence}, where a
remarkable enhancement of the emission intensity is found as
compared to the case of an isolated quantum well in free space.
Conclusions and perspectives are finally drawn in Sec.
\ref{conclusions}.

\section{Model Hamiltonian}
\label{introduction}

\begin{figure}[t!]
\begin{center}
\includegraphics[width= 9 cm]{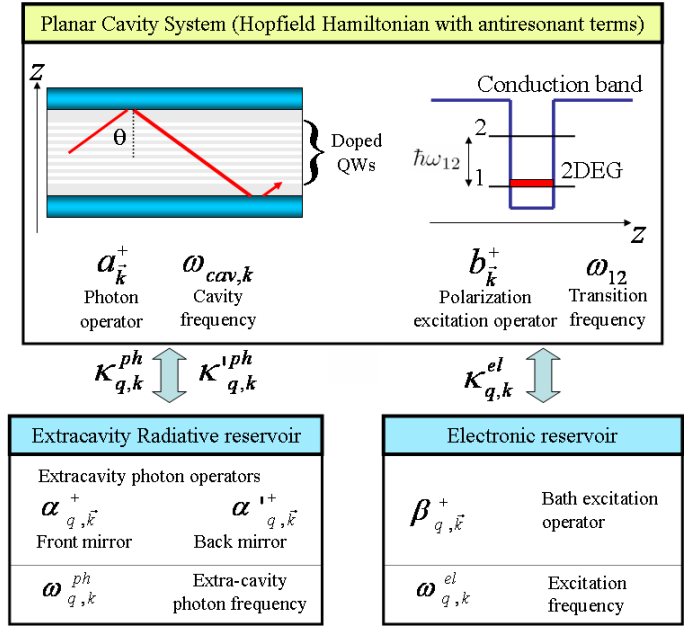}
 \caption{\label{sketch} A sketch of the investigated model with the
 corresponding operators
 and frequencies. The system is represented by a planar cavity mode embedding
 a doped multiple quantum well structure. The cavity photon field is coupled to
 the quantum well electronic polarization associated to a resonant intersubband transition
 in the doped quantum wells. The Hopfield Hamiltonian for the cavity
 system includes the anti-resonant
 light-matter interaction terms, which are significant
 in the ultra-strong coupling regime\cite{Ciuti_vacuum} (vacuum Rabi
 frequency $\Omega_{R,k}$ comparable
 to the transition frequency $\omega_{12}$).
The cavity mode is coupled to the extracavity
 electromagnetic field. The electronic polarization is coupled to
 a bath of electronic excitations.}
\end{center}
\end{figure}

As schematically shown in Fig.\ref{sketch}, the Hamiltonian of the present system
contains three main blocks:
\begin{equation}
H= \label{eq:Hamilt} H_{sys} + H^{phot}_{bath} + H^{el}_{bath}~,
\end{equation}
where $H_{sys}$ is the Hamiltonian
describing the closed cavity system, while $H^{phot}_{bath}$ and
$H^{el}_{bath}$ take into account the coupling to respectively the
photonic and electronic reservoirs.

The Hamiltonian $H_{sys} = H_0+H_{res}+H_{anti}$ for the closed
cavity system has the typical Hopfield~\cite{Hopfield} form
already discussed in detail in~Ref. \onlinecite{Ciuti_vacuum}, and
contains three parts respectively describing the energy of the
bare cavity photons and of the electronic excitations ($H_0$), the
resonant part of the light-matter interaction ($H_{res}$), and the
anti-resonant terms \footnote{Note that the anti-resonant terms
$H_{anti}$ in the
  Hamiltonian are due to the same {\it vacuum} Rabi coupling
  responsible for the resonant terms $H_{res}$.
Hence, this problem, although there is some formal reminiscence, is different
from the case of antiresonant transitions and Bloch-Siegert-like shifts \cite{BlochSiegert}
induced by a {\it classical} Rabi coupling produced by an intense and coherent pump field,
which is comprehensively studied in Ref.\onlinecite{Lewenstein}.}
usually neglected in the so-called rotating-wave
  approximation~\cite{CCT4} ($H_{anti}$):
\begin{eqnarray}
H_{0} & =&  \sum_{\bm{k}}  \hbar \omega_{{\rm cav},k} \left (
a^{\dagger}_{\bm{k}} a_{\bm{k}} + \frac{1}{2} \right ) + \sum_{\bm
k} \hbar \omega_{12}~b^{\dagger}_{\bm{k}} b_{\bm{k}}   ~,~
\label{H_0} \\
H_{res} &=& \hbar \sum_{\bm{k}} \{ i \Omega_{R,k}  (
a^{\dagger}_{\bm{k}} b_{\bm{k}} - a_{\bm{k}} b^{\dagger}_{\bm{k}}
)   + D_k ( a^{\dagger}_{\bm{k}} a_{\bm{k}} + a_{\bm{k}}
a^{\dagger}_{\bm{ k}} )
 \}~,~ \label{H_res} \\
H_{anti} &=& \hbar \sum_{\bm{k}}  \{ i \Omega_{R,k}  ( a_{\bm{k}}
b_{\bm{-k}} - a^{\dagger}_{\bm{k}} b^{\dagger}_{- \bm{k}}  )
 +  D_k ( a_{\bm{k}} a_{\bm{-k}}
+ a^{\dagger}_{\bm{k}} a^{\dagger}_{\bm{- k}}) \}~. \label{H_anti}
\end{eqnarray}
Translational invariance along the cavity plane has been here assumed.
$a^{\dagger}_{\bf k}$ is the creation operator of a cavity photon with
in-plane wave-vector ${\bf k}$ and energy $\hbar \omega_{{\rm
cav},\kk}$.
$b^{\dagger}_{\bf k}$ is the creation operator of
the  {\em bright} intersubband excitation mode of wavevector ${\bf
    k}$ of the doped quantum well~\cite{Ciuti_vacuum}:
\begin{equation}
b^{\dagger}_{\bf
k}=\frac{1}{\sqrt{N_{QW}\,\sigma_{el}\,S}}\sum_{j=1}^{N_{QW}}\,\sum_{|{\bf
q}|<k_F} c^{(j) \dagger}_{2,{\bf
      q}+{\bf k}}\,c^{(j)}_{1,{\bf q}}.
\label{b_dagger}
\end{equation}
Here, $N_{QW}$ is the number of quantum wells present in the
cavity (which are assumed for simplicity to be situated at the
antinodes of the cavity mode, and therefore identically coupled to
the photonic mode), $\sigma_{el} = N_{el}/S$ is electron density
per unit area in each quantum well and $S$ is the quantization
area. The fermionic operator $c^{(j)}_{1,{\bf q}}$ annihilates an
electron of in-plane wavevector ${\bf q}$ from the lowest subband
of the $j$-th quantum well, while $c^{(j)\,\dagger}_{2,{\bf q}}$
creates an electron with wavevector ${\bf q}$ in the second,
excited, subband of the $j$-th quantum well. $k_F$ is the Fermi
wavevector of the two-dimensional electron gas in each well, and
the electronic ground state is written as:
\begin{equation}
|F\rangle=\prod_{j=1}^{N_{QW}}\,\prod_{|{\bf q}|<k_F} c^{(j)\,\dagger}_{1,{\bf q}}\,|0_{cond}\rangle
\label{el_ground}
\end{equation}
in terms of the empty conduction band state $|0_{cond}\rangle$.
In the following, we shall always restrict ourselves to a weak
excitation regime:
\begin{equation}
\frac{1}{S}\sum_{{\bf k}} \big\langle b_{{\bf k}}^\dagger\,b_{{\bf
    k}}\,\big\rangle \ll \sigma_{el},
\end{equation}
in which the operators $b_{{\bf k}}$ are approximately
bosonic:
\begin{equation}
\big[b_{{\bf k}},b_{{\bf k}'}^\dagger \big]=\delta_{{\bf k},{\bf k}'}.
\end{equation}
The $N_{QW}N_{el}-1$ states orthogonal to $b^{\dagger}_{\kk} |F\rangle$ are
dark excitations (uncoupled to the cavity photon) which do not play any role in the
  optical processes taking place in the system.
The first and second electronic subbands have been here assumed to
  be perfectly parallel~\cite{bastard}, with a momentum-independent transition
  frequency $\omega_{12}$.

$\Omega_{R,k}$ is the vacuum Rabi frequency which quantifies the
strength of the light-matter dipole coupling: in the case of intersubband
transitions in doped quantum wells \cite{Ciuti_vacuum}, it can become a
significant fraction of the intersubband transition $\omega_{12}$. The
explicit expression is
\begin{equation}
\Omega_{R,k} = \left ( \frac{2 \pi e^2}{\epsilon_{\infty}m _0
L^{\rm eff}_{\rm cav}} ~\sigma_{el} ~N_{QW}~f_{12} \sin^2
\theta \right  )^{1/2}~, \label{Rabi}~
\end{equation}
where $L^{\rm eff}_{\rm cav}$ is the effective length of the
cavity mode and $\theta$ is the intracavity photon propagation
angle such that $\sin\theta=c k /(\omega_{12}
\sqrt{\epsilon_{QW}})$.

$D_k$ is the related term originating from the squared electromagnetic
vector potential term in the minimal coupling light-matter
Hamiltonian~\cite{Ciuti_vacuum}: for a typical quantum well potential,
one has $D_k \simeq \Omega_{R,k}^2/\omega_{12}$.

The environment of the open cavity system is modeled by two baths
of excitations, associated to respectively the photonic and the
  electronic degrees of freedom.
In this paper, we shall focus our attention on cavity
configurations in which the photonic mode is coupled to the
external electromagnetic field via the finite transmittivity of
the planar mirrors enclosing the cavity, as recently done
\cite{Dini_PRL,Liu,Aji_APL}. In a planar geometry, the coupling of
the cavity photon to the extra-cavity electromagnetic modes is
well described by the Hamiltonian:
\begin{equation}
\label{ph_bath} H^{ph}_{bath} = \int dq \sum_{\bf k} \hbar
\omega^{ph}_{{q},{\bf k}} \left(\alpha^{\dagger}_{{q},\bf k}
\alpha_{{q},\bf k} + \frac{1}{2} \right)+ i \hbar \int dq
\sum_{\bf k} \left(\kappa^{ph}_{{q},{\bf k}}\,\alpha_{{q},{\bf k}}
~a^{\dagger}_{\bf k}-  \kappa^{ph\,*}_{{q},{\bf
k}}\,\alpha^{\dagger}_{{q},{\bf k}} a_{\bf k} \right)~,
\end{equation}
where $\omega^{ph}_{{q},{\bf k}}$ is the frequency of an
extra-cavity photon with in-plane wavevector ${\bf k}$ and
wavevector ${q}$ in the orthogonal direction and
$\alpha^{\dagger}_{{q},\bf k}$ is the corresponding creation
operator, obeying the commutation rule $[\alpha_{{q},\bf
k},\alpha^{\dagger}_{{q'},{\bf k'}}] = \delta(q-q') \delta_{{\bf
k},{\bf k'}}$. The coupling between the cavity and the extracavity
radiation fields is quantified by the tunneling matrix element
$\kappa^{ph}_{{q},{\bf k}}$ through the cavity mirror, whose value
depends on the specific mirror structure and can be calculated by
solving the classical Maxwell equations.

The Hamiltonian (\ref{ph_bath}) takes into account a single radiative
bath, coupled to the cavity through one of its mirrors.
For the sake of simplicity, we shall start our analysis from this
simplified case, which is then extended in Sec.\ref{double_section} to
the more realistic case of two photonic baths coupled to the cavity
through the front and the back mirrors.
All this discussion is performed by neglecting the anti-resonant terms in
$H^{phot}_{bath}$ whose effect is quantitatively small; the general
case including would not however pose any additional problem,
and is considered and solved in Appendix \ref{General}.

Concerning the bath coupled to the electronic transition, the
modelling is not as straightforward as in the photonic case.
In the case of intersubband transitions in doped
semiconductor quantum wells\cite{ISB-Book,ISB-Book_2}, the damping and decoherence of the
electronic transition at the frequency $\omega_{12}$ is due to the
interplay of different processes, such as the interaction with
crystal phonons (optical and acoustical), the scattering with impurities and with free
carriers in levels other from the ones involved in the considered electronic
transition\cite{Robson}.
As our purpose is to study the role of the electronic losses
and decoherence on the peculiar quantum optical properties of
cavities in the ultra-strong coupling regime, we do not attempt here
to address the specific nature of the decoherence channels, but we
will rather model them as a phenomenological bath of harmonic
excitations in the spirit of the Caldeira-Leggett
model~\cite{Caldeira-Leggett}.
As for the photonic bath, the harmonic oscillators modes coupled to
the intersubband excitation of wavevector ${\bf k}$ are labelled
by a continuous index $q$ and have frequency $\omega^{el}_{{q},{\bf
    k}}$.
They affect the electronic intersubband transition according to the Hamiltonian:
\begin{equation}
H^{el}_{bath} = \int dq \sum_{\bf k} \hbar \omega^{el}_{{q},{\bf
k}} \left(\beta^{\dagger}_{{q},\bf k} \beta_{{q},\bf k}  +
\frac{1}{2} \right)
\nonumber \\
+ i \hbar  \int dq \sum_{\bf k} \left[ \kappa^{el}_{{q},{\bf
k}}\,\beta_{{q},{\bf k}} ~b^{\dagger}_{\bf k}-
\kappa^{el\,*}_{{q},{\bf
    k}}\,\beta^{\dagger}_{{q},{\bf k}} b_{\bf
k}\right]~.
\label{el_bath}
\end{equation}
Here, the bath operators $\beta_{{q},\bf k}$ satisfy the harmonic
oscillator commutation rule $[\beta_{{q},\bf
k},\beta^{\dagger}_{{q'},{\bf k'}}] = \delta(q-q') \delta_{{\bf
k},{\bf k'}}$. $\kappa^{el}_{{q},{\bf k}}$ are the matrix elements
quantifying the coupling to the electronic polarization. Inclusion
of the anti-resonant terms in $H^{el}_{bath}$ is straightforward,
as discussed in Appendix \ref{General}. In practice,
$\omega^{el}_{{q},{\bf k}}$ and $\kappa^{el}_{{q},{\bf
    k}}$ can be chosen at will so as to quantitatively reproduce the
properties of a specific system, in particular the
frequency-dependent damping rate of the electronic excitations.
A recapitulative scheme of the investigated model in depicted in
Fig. \ref{sketch}. For a more detailed description of the specific
system of semiconductor microcavities embedding intersubband
transitions, the reader can refer to Refs. \onlinecite{Dini_PRL}
and \onlinecite{Ciuti_vacuum}.
\section{Quantum dynamical equations}
\label{Langevin}

\subsection{Quantum Langevin equations}
The equation of motion for the extra-cavity photon operator in
Heisenberg representation reads:
\onecolumngrid
\begin{equation}
\frac{d\alpha_{{q},{\bf k}}}{dt} =
-\frac{i}{\hbar} [\alpha_{{q},{\bf k}},H] = - i
\omega^{ph}_{{q},{\bf k}} \alpha_{{q},{\bf k}} -
\kappa^{ph\,*}_{{q},{\bf k}} a_{\bf k} ~,
\end{equation}
and its solution can be formally written as
\begin{equation}
\alpha_{{q},{\bf k}}(t) =
e^{-i\omega^{ph}_{{q},{\bf k}} (t-t_0)} \alpha_{{q},{\bf k}}(t_0)
- \kappa^{ph\,*}_{{q},{\bf k}} \int_{t_0}^{t} dt'
e^{-i\omega^{ph}_{{q},{\bf k}} (t-t')} a_{\bf k}(t') ~,
\label{formal_1}
\end{equation}
$t_0$ being the initial time.
Inserting these formulas into the evolution equation
for the cavity photon amplitude, one finds:
\begin{multline}
\frac{da_\kk}{dt}=-\frac{i}{\hbar}[a_\kk,H_{sys}]+ \int dq~
k^{ph}_{{q},\kk}\,\alpha_{{q},\kk}= \\
=-\frac{i}{\hbar}[a_\kk,H_{sys}]+ \int dq~
k^{ph}_{{q},\kk}\,\alpha_{{q},\kk}(t_0)\,e^{-i\omega^{ph}_{{q},\kk}(t-t_0)}
-\int
dq~|k_{{q},\kk}|^2\,\int_{t_0}^tdt'\,e^{-i\omega^{ph}_{{q},\kk}(t-t')}\,
a_\kk(t') \label{da/dt}~.
\end{multline}

%\twocolumngrid

Using the standard definition
\begin{equation}
\alpha^{in}_{{q},\kk}=\alpha_{{q},\kk}(t_0)\,e^{-i\omega^{ph}_{{q},\kk} t_0}
\label{a_input}
\end{equation}
for the input fields at $t_0\rightarrow -\infty$, one can cast
(\ref{da/dt}) in the
form of a quantum Langevin equation:
\begin{equation}
\label{eq_a} \frac{d a_{\bf k}}{d t} = -
\frac{i}{\hbar} [a_{\bf k},H_{sys}]  - \int_{-\infty}^{\infty} dt'
\,\Gamma_{{\rm cav},\bf k}(t-t') ~a_{\bf k}(t')  + F_{{\rm cav}, \bf k}(t)~,
\end{equation}
where the (causal) damping memory kernel is given by
\begin{equation}
\Gamma_{{\rm cav},\bf k}(\tau) = \Theta(\tau)~\int dq~
|\kappa^{ph}_{{q},{\bf k}}|^2\, e^{-i\omega^{ph}_{{q},{\bf k}}
\tau}~,
\end{equation}
and the fluctuating Langevin force is represented by the operator
\begin{equation}
F_{{\rm cav},\bf k}(t) = \int dq~ \kappa^{ph}_{{q},{\bf k}}
e^{-i\omega^{ph}_{{q},{\bf k}} t} \alpha^{in}_{{q},{\bf k}}~.
\label{F_cav}
\end{equation}
Analogously, one obtains a quantum Langevin equation for the
electronic polarization field
\begin{equation}
\label{eq_b} \frac{\partial b_{\bf k}}{\partial t} = -
\frac{i}{\hbar} [b_{\bf k},H_{sys}]  - \int_{t_0}^{\infty} dt'
\Gamma_{12,\bf k}(t-t') ~b_{\bf k}(t')  + F_{12, \bf k}(t)~,
\end{equation}
in terms of the memory kernel
\begin{equation}
\Gamma_{12,\bf k}(\tau) = \Theta(\tau)~\int dq~
|\kappa^{el}_{{q},{\bf k}}|^2 e^{-i\omega^{el}_{{q},{\bf k}}
\tau}~,
\end{equation}
and the Langevin force
\begin{equation}
F_{12,\bf k}(t) = \int dq~ \kappa^{el}_{{q},{\bf k}}
e^{-i\omega^{el}_{{q},{\bf k}} t} \beta^{in}_{{q},{\bf k}}.
\end{equation}
The input operators $\beta^{in}_{{q},\kk}$ for the electronic bath
are here defined in the same way as the photonic ones (\ref{a_input}).

\subsection{Input-output relations}

The extra-cavity asymptotic {\em output} operators at $t = +\infty$
can be related to the {\em input} operators at $t_0 = - \infty$ and the
cavity photon ones through a linear relationship. Taking $t_0
\to -\infty$ and $t\to +\infty$ in Eq. (\ref{formal_1}), we obtain the formula
\begin{equation}
\label{input-output_a} \alpha^{out}_{{q},{\bf k}} =
\alpha^{in}_{{q},{\bf k}}  - \kappa^{ph\star}_{{q},{\bf k}}~
\tilde{a}_{\bf k}(\omega^{ph}_{{q},{\bf k}}),
\end{equation}
where $\tilde{a}_{\bf k}(\omega)$ is the Fourier transform\cite{note_FFT} of
$a_{\bf k}(t)$.
An analogous expression holds for the electronic bath operators:
\begin{equation}
\label{input-output_b} \beta^{out}_{{q},{\bf k}} =
\beta^{in}_{{q},{\bf k}}  - \kappa^{el\star}_{{q},{\bf k}}~
\tilde{b}_{\bf k}(\omega^{el}_{{q},{\bf k}})
\end{equation}
\begin{figure}[t!]
\begin{center}
\includegraphics[width= 9 cm]{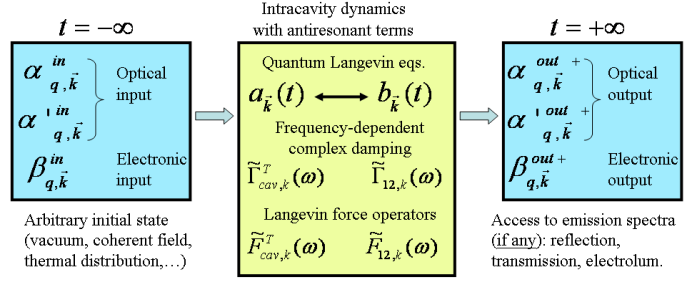}
 \caption{\label{fig:input} A recapitulative sketch of the
 input-output scheme.
 The initial condition on the extracavity photon operators and the
 electronic bath operators at $t=-\infty$ represents the {\em input}.
 The intracavity dynamics
 for the cavity mode and electronic polarization field is described by
 quantum Langevin equations including
 the antiresonant terms of the light-matter interaction. The radiative
 and non-radiative baths produce a complex frequency-dependent damping
 as well as frequency-dependent  Langevin forces.
The bath operators at $t= + \infty$ represent the radiative and
 non-radiative {\em output}.}
\end{center}
\end{figure}

\subsection{Equations in frequency space}
In the present case of a time-independent $H_{sys}$, the quantum
Langevin equations (\ref{eq_a}) and (\ref{eq_b}) are most conveniently
solved in the frequency space. By Fourier transforming them and their
hermitian conjugates\cite{note_FFT}, we get the following equation
 \onecolumngrid
\begin{equation}
\label{dyn_eq} \bar{\mathcal M}_{{\bf k},\omega} ~ \left (
\begin{array}{c}
\tilde{a}_{\bf k}(\omega) \\
\tilde{b}_{\bf k}(\omega) \\
\tilde{a}^{\dagger}_{-\bf k}(-\omega) \\
\tilde{b}^{\dagger}_{-\bf k}(-\omega) \\
\end{array}
\right)  + i  \left (
\begin{array}{c}
\tilde{F}_{{\rm cav},{\bf k}}(\omega) \\
\tilde{F}_{12,{\bf k}}(\omega) \\
\tilde{F}^{\dagger}_{{\rm cav},{-\bf k}}(-\omega) \\
\tilde{F}^{\dagger}_{12,{-\bf k}}(-\omega) \\
\end{array}
\right) = 0 ~,
\end{equation}
where the 4$\times$4 matrix
\begin{equation}
\label{Hopfield_omega}
\bar{\mathcal M}_{{\bf k},\omega} =\left (
\begin{array}{cccc}
\omega_{{\rm cav},k} + 2 D_k-\omega - i \tilde{\Gamma}_{{\rm cav},{\bf k}}(\omega)  & i \Omega_{R,k} & 2D_k & -i \Omega_{R,k} \\
-i\Omega_{R,k} & \omega_{12} -\omega - i \tilde{\Gamma}_{12,{\bf k}}(\omega) & - i \Omega_{R,k} & 0 \\
-2 D_k & -i \Omega_{R,k} & -\omega_{{\rm cav},k} - 2 D_k -\omega - i \tilde{\Gamma}^{\star}_{{\rm cav},{\bf k}}(-\omega) & i \Omega_{R,k} \\
-i \Omega_{R,k} &  0 & -i \Omega_{R,k} &  -\omega_{12} -\omega - i \tilde{\Gamma}^{\star}_{12,{\bf k}}(-\omega)\\
\end{array}
\right )
\end{equation}
%\twocolumngrid
is the Hopfield~\cite{Hopfield,Ciuti_vacuum} matrix of our system,
additioned by the terms accounting for the damping and for the
frequency shift produced by the coupling to the baths.
The complex self-energy shift of the cavity photon due to the coupling
to the photonic bath is indeed
\begin{equation}
\tilde{\Gamma}_{{\rm cav},\bf k}(\omega) = \int dq~
\pi\,|\kappa^{ph}_{{q},{\bf k}}|^2
\delta(\omega-\omega^{ph}_{{q},{\bf k}}) + i {\mathcal P} \int dq~
\frac{|\kappa^{ph}_{{q},{\bf k}}|^2}{\omega-\omega_{{q},{\bf k}}}.
\label{Gamma_om}
\end{equation}
Its real part  represents the frequency-dependent radiative damping
of the cavity mode due to radiative losses. This, in agreement with
the usual Fermi golden rule
\begin{equation}
\Re[{\tilde \Gamma}_{{\rm cav},\kk}(\omega)]=
\pi\,|k_{\bar{q},\kk}^{ph}|^2\,\rho^{ph}_\kk(\omega),
\end{equation}
where $\rho_\kk^{ph}(\omega)=\big[d\omega_{q =
 \bar{q},\kk}^{ph}/dq\big]^{-1}$ is
 the photonic density of states at in-plane wavevector $\kk$ and $\bar{q}$ is the
 resonant wavevector such that $\omega_{\bar{q},\kk}^{ph}=\omega$.
The imaginary part $\Im[{\tilde
    \Gamma}_{{\rm cav},\kk}(\omega)]$ accounts instead for the corresponding
 Lamb shift~\cite{CCT4}: for a given $\Re[{\tilde
 \Gamma}_{{\rm cav},\kk}(\omega)]$, the Lamb shift $\Im[{\tilde
    \Gamma}_{{\rm cav},\kk}(\omega)]$ is in fact univocally fixed by the
Kramers-Kronig causality relationships
\begin{equation}
\label{Kramers}
\Im \{ \tilde{\Gamma}_{{\rm cav},{\bf k}}(\omega) \} = - \frac{1}{\pi}
{\mathcal P}~\int_{-\infty}^{\infty} d \omega' \frac{\Re \left \{
\tilde{\Gamma}_{{\rm cav},{\bf k}}(\omega')\right \}}{\omega'-\omega}~.
\end{equation}

An analogous formula holds for the electronic excitation counterpart:
\begin{equation}
\tilde{\Gamma}_{12,\bf k}(\omega) = \int dq~
|\kappa^{el}_{{\bar{q}},{\bf k}}|^2 \pi
\delta(\omega-\omega^{el}_{{q},{\bf k}}) + i {\mathcal P} \int dq~
\frac{|\kappa^{el}_{{q},{\bf k}}|^2}{\omega-\omega^{el}_{{q},{\bf
k}}}~,
\end{equation}
The real part of $\tilde{\Gamma}_{12,\bf k}(\omega)$ represents
the frequency-dependent broadening of the electronic transition, of
non-radiative origin.
As all real excitations of the considered baths have by definition
positive frequency $\omega^{ph,el}_{{q},{\bf k}}> 0$, one has:
\begin{equation}
\Re \{ \tilde{\Gamma}_{{\rm cav},\bf k}(\omega < 0) \} =
\Re \{\tilde{\Gamma}_{12,\bf k}(\omega < 0) \} = 0~.
\label{zero_1}
\end{equation}
In frequency space, the Langevin forces have the form
\begin{eqnarray}
\tilde{F}_{{\rm cav},\bf k}(\omega) &=& \int dq~ \kappa^{ph}_{{q},{\bf
k}}~ 2 \pi \delta(\omega-\omega^{ph}_{{q},{\bf
k}})~\alpha^{in}_{{q},{\bf k}}=2\pi
\kappa^{ph}_{\bar{q}}\rho_\kk^{ph}(\omega)\,\alpha^{in}_{{\bar q},\kk}
\label{Fcav} \\
\tilde{F}_{12,\bf k}(\omega) &= &\int dq'~ \kappa^{el}_{{q'},{\bf
k}}~ 2 \pi \delta(\omega-\omega^{el}_{{q'},{\bf
k}})~\beta^{in}_{{q'},{\bf k}}=2\pi
\kappa^{el}_{\bar{q}'}\rho_\kk^{el}(\omega)\,\beta^{in}_{{\bar
q}',\kk}, \label{F12}
\end{eqnarray}
where ${\bar q}$ and ${\bar q}'$ in the right-hand sides are such that
$\omega^{ph}_{{\bar q},\kk} = \omega^{el}_{{\bar q}',\kk}=\omega$.
In analogy with (\ref{zero_1}), one therefore has
\begin{equation}
\tilde{F}_{{\rm cav},\bf k}(\omega < 0)  =  \tilde{F}_{12,\bf k}(\omega < 0)
= 0~.
\end{equation}
A recapitulative sketch of the input-output framework here developed is drawn
in Fig. \ref{fig:input}.

\section{Exact solutions for the operators}
\label{solutions}
The equation (\ref{dyn_eq}) for the Fourier space cavity field
operators is immediately solved by matrix inversion:
\begin{equation}
  \left (
\begin{array}{c}
\tilde{a}_{\bf k}(\omega) \\
\tilde{b}_{\bf k}(\omega) \\
\tilde{a}^{\dagger}_{-\bf k}(-\omega) \\
\tilde{b}^{\dagger}_{-\bf k}(-\omega) \\
\end{array}
\right) =  \bar{\mathcal G}({{\bf k},\omega})  \left (
\begin{array}{c}
\tilde{F}_{{\rm cav},{\bf k}}(\omega) \\
\tilde{F}_{12,{\bf k}}(\omega) \\
\tilde{F}^{\dagger}_{{\rm cav},{-\bf k}}(-\omega) \\
\tilde{F}^{\dagger}_{12,{-\bf k}}(-\omega) \\
\end{array}
\right)
\label{dyn_eq2}
\end{equation}
where
\begin{equation}
\bar{\mathcal G}({{\bf k},\omega}) = -i~[\bar{\mathcal M}_{{\bf
k},\omega}]^{-1}~.
\end{equation}
For $\omega > 0$, this can be simplified as
\begin{eqnarray}
\tilde{a}_{\bf k}(\omega > 0) & =&  \bar{\mathcal G}_{11}({\bf
k},\omega)~ \tilde{F}_{{\rm cav},{\bf k}}(\omega )
          + \bar{\mathcal G}_{12}({{\bf k},\omega})
           ~ \tilde{F}_{12,{\bf k}}(\omega)~, \label{a} \\
\tilde{b}_{\bf k}(\omega > 0) & =& \bar{\mathcal G}_{21}({\bf
k},\omega)~ \tilde{F}_{{\rm cav},{\bf k}}(\omega )
          + \bar{\mathcal G}_{22}({{\bf k},\omega})
           ~ \tilde{F}_{12,{\bf k}}(\omega). \label{b}
\end{eqnarray}
Using the expressions Eqs. (\ref{input-output_a}-\ref{input-output_b}) for the output
field and the expressions (\ref{Fcav}-\ref{F12}) for the Langevin
forces, one finally gets to the {\em input-output} relation:
\begin{equation}
  \left (
\begin{array}{c}
\alpha^{out}_{{q},{\bf k}} \\
\beta^{out}_{{q}',{\bf k}} \\
\end{array}
\right) = \left (
\begin{array}{cc}
\bar{\mathcal U}_{11}({{\bf k},\omega}) &  \bar{\mathcal U}_{12}({{\bf k},\omega})\\
\bar{\mathcal U}_{21}({{\bf k},\omega}) &  \bar{\mathcal U}_{22}({{\bf k},\omega})\\
\end{array}
\right)
   \left (
\begin{array}{c}
\alpha^{in}_{{q},{\bf k}} \\
\beta^{in}_{{q}',{\bf k}} \\
\end{array}
\right). \label{unitary}
\end{equation}
This is a linear relation linking the output operators of the photonic
and electronic bath modes to the corresponding input ones; as expected
by the time-invariance of the Hamiltonian under examination, the
pair of photonic and electronic modes involved in (\ref{unitary})
share by construction the same frequency
$\omega^{ph}_{q,\kk}=\omega^{el}_{q',\kk}$.
The matrix elements of ${\mathcal U}$ are
\begin{eqnarray}
\bar{\mathcal U}_{11}({{\bf k},\omega}) =  1-  2 \Re
\{\tilde{\Gamma}_{{{\rm cav}},\bf k}(\omega)\} \bar{\mathcal
G}_{11}({{\bf k},\omega}) ~, \label{U11} \\
\bar{\mathcal U}_{12}({{\bf k},\omega}) =  -  2 \Re
\{\tilde{\Gamma}_{{12},\bf k}(\omega)\}
\frac{\kappa^{ph\,*}_{{q},{\bf k}}}{\kappa^{el\,*}_{{q}',{\bf
k}}} \bar{\mathcal G}_{12}({{\bf k},\omega}) ~, \label{U12}\\
          \bar{\mathcal U}_{21}({{\bf k},\omega}) =  -  2 \Re
\{\tilde{\Gamma}_{{{\rm cav}},\bf k}(\omega)\}
\frac{\kappa^{el\,*}_{{q}',{\bf
k}}}{\kappa^{ph\,*}_{{q},{\bf k}}}  \bar{\mathcal G}_{21}({{\bf
    k},\omega}) ~, \label{U21}\\
          \bar{\mathcal U}_{22}({{\bf k},\omega}) =  1-  2 \Re
\{\tilde{\Gamma}_{{12},\bf k}(\omega)\} \bar{\mathcal
G}_{22}({{\bf k},\omega}) ~.\label{U22}
\end{eqnarray}
The unitarity of the matrix $\bar{\mathcal U}({{\bf k},\omega})$
(that has been numerically checked) guarantees that the usual Bose
commutation rules hold for the output operators
$\alpha^{out}_{{q},\kk}$ and $\beta^{out}_{{q},\kk}$, namely
$[\alpha^{out}_{q,\kk},\alpha^{out\,\dagger}_{q',\kk'}]=\delta(q-q')\,\delta_{\kk,\kk'}$
and
$[\beta^{out}_{q,\kk},\beta^{out\,\dagger}_{q',\kk'}]=\delta(q-q')\,\delta_{\kk,\kk'}$.
>From a more physical point of view, the unitarity of $\bar{\mathcal
  U}({{\bf k},\omega})$ means that the total energy is conserved
during a scattering process when energy is sent onto the cavity as
radiation or as electronic energy, and both the emerging light and
the electronic absorption are taken into account:
\begin{equation}
\left\langle \alpha^{out\,\dagger}_{q,\kk}\,\alpha^{out}_{q,\kk}+
\beta^{out\,\dagger}_{q',\kk}\,\beta^{out}_{q',\kk}   \right\rangle=
\left\langle \alpha^{in\,\dagger}_{q,\kk}\,\alpha^{in}_{q,\kk}+
\beta^{in\,\dagger}_{q',\kk}\,\beta^{in}_{q',\kk}   \right\rangle
\end{equation}

It is interesting and an important check of consistency to verify
that the usual commutation rules
$[a_\kk(t),a^\dagger_{\kk'}(t)]=\delta_{\kk,\kk'}$ hold for the
cavity (as well as for the electronic polarization) operators.
This has been verified on our specific model in the following way.
By definition, one has:
\begin{equation}
[a_\kk(t),a^\dagger_{\kk'}(t)]=\int\frac{d\omega}{2\pi}\int\frac{d\omega'}{2\pi}\,e^{-i\omega
  t}\,e^{i\omega' t}\,[{\tilde a}_\kk(\omega),{\tilde a}^\dagger_{\kk'}(\omega')].
\end{equation}
 Inserting here (\ref{dyn_eq2}), and noticing that
\begin{equation}
[{\tilde F}_{{\rm cav},\kk}(\omega),{\tilde F}_{{\rm cav},\kk'}^\dagger(\omega')]=2\pi\,\delta(\omega-\omega')\,2\,\Re[{\tilde
    \Gamma}_{{\rm cav},\kk}(\omega)]\,\delta_{\kk,\kk'},
\end{equation}
one can finally write:
\begin{multline}
[a(t),a^\dagger(t)]=2\int\!\frac{d\omega}{2\pi}\, \Big[ |\bar{\mathcal
G}_{11}({\bf k},\omega)|^2\,\Re[{\tilde
    \Gamma}_{{\rm cav},\kk}({\bf k},\omega)]+|\bar{\mathcal G}_{12}({\bf k},\omega)|^2\,\Re[{\tilde
    \Gamma}_{12,\kk}(\omega)]+ \\
- |\bar{\mathcal G}_{13}({\bf k},\omega)|^2\,\Re[{\tilde
    \Gamma}_{{\rm cav},\kk}(-\omega)]-
|\bar{\mathcal G}_{14}({\bf k},\omega)|^2\,\Re[{\tilde
    \Gamma}_{12,\kk}(-\omega)]
 \Big],
\end{multline}
which has been numerically checked to give $1$ as expected. This
fact critically depends on the consistent inclusion of the real and
imaginary parts of ${\tilde \Gamma}_{{\rm cav},\kk}(\omega)$ and ${\tilde
  \Gamma}_{12,\kk}(\omega)$ satisfying the
causality relation (\ref{Kramers}).

Note finally that the presence of the anti-resonant terms in $H_{sys}$
implies that the
cavity operators ${\tilde a}_\kk(\omega)$ and ${\tilde b}_\kk(\omega)$
have non-vanishing values also for negative frequencies:
\begin{eqnarray}
\tilde{a}_{\bf k}(\omega < 0) & =& \bar{\mathcal G}_{13}({\bf
k},\omega)~ \tilde{F}^{\dagger}_{{\rm cav},{-\bf k}}(-\omega )
          + \bar{\mathcal G}_{14}({{\bf k},\omega})
           ~ \tilde{F}^{\dagger}_{12,{-\bf k}}(-\omega) \\
\tilde{b}_{\bf k}(\omega < 0) & =&  \bar{\mathcal G}_{23}({\bf
k},\omega)~ \tilde{F}^{\dagger}_{{\rm cav},{-\bf k}}(-\omega )
          + \bar{\mathcal G}_{24}({{\bf k},\omega})
           ~ \tilde{F}^{\dagger}_{12,{-\bf k}}(-\omega)~.
\end{eqnarray}

\section{The case of a vacuum input: properties of the intracavity quantum ground state}
\label{ground}

\begin{figure}[t!]
\begin{center}
\includegraphics[width= 9 cm]{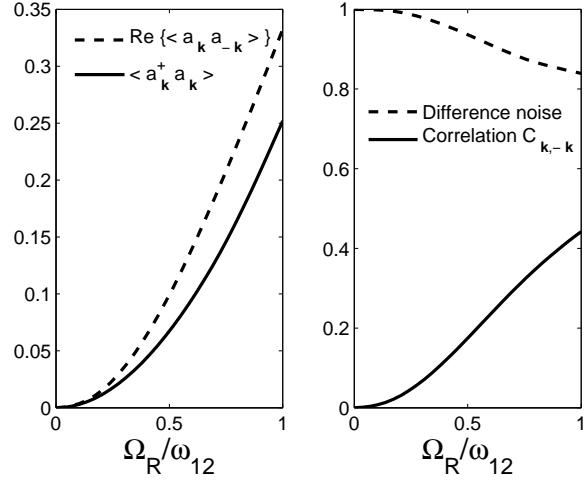}
\caption{\label{fig_groundstate} Quantum properties of the
 intra-cavity ground state for input fields in the
 vacuum state.
 In this case, the output is also in the vacuum state, which
 implies that the cavity photons present in the cavity are purely
 virtual ones and can not escape the cavity.
 Left panel: cavity photon number $\langle a^{\dagger}_{\bf k}
 a_{\bf k} \rangle$ (solid line) and the real part of the anomalous expectation value
 $\langle a_{\bf k} a_{-\bf k} \rangle$ (dashed line) as a
 function of the normalized vacuum Rabi frequency
 $\Omega_R/\omega_{12}$.
Right panel: corresponding cavity photon correlation $C_{{\bf
k},-{\bf k}}$ (solid line) and the normalized difference noise $F_{{\rm diff}}$ (dashed line).
$\omega_{{\rm cav},k} = \omega_{12}$, and the damping rates are taken as
 frequency-independent ones. The calculations have been performed with
 $\Re\{\Gamma_{{\rm cav},\kk}(\omega > 0)\}= \bar{\Gamma}_{{\rm cav},k}=
\Re\{\Gamma'_{{\rm cav},\kk}(\omega > 0)\}= \bar{\Gamma}'_{{\rm cav},k}=
0.04  \omega_{12}$
and $\Re \{ \Gamma_{12,k}(\omega >0) \} = \bar{\Gamma}_{12,k}= 0.08\omega_{12}$.}
\end{center}
\end{figure}
Once we have obtained explicit expressions for the output
operators in terms of the input ones, we can apply this formalism
to study the quantum dissipative response of the cavity system for
different initial states.

We shall start from the simplest case of a bath initially in its
vacuum state $|0\rangle$ such that:
\begin{equation}
\alpha^{in}_{{q},{\bf k}} |0 \rangle = \beta^{in}_{{q},{\bf k}} |0
\rangle = 0
\end{equation}
for every ${\bf k}$ and $q$. As expected on physical
grounds, no excitations are created, and the bath remains in its
vacuum state. It is in fact a straightforward consequence of
(\ref{unitary}) that
\begin{equation}
\alpha^{out}_{{q},{\bf k}} |0\rangle = \beta^{out}_{{q},{\bf k}}
|0\rangle = 0.
\end{equation}
On the other hand, the cavity system itself is not in the usual vacuum
state, and a finite number of photons is present in this anomalous ground state.
This can be calculated by means of (\ref{dyn_eq2}), which relates the
in-cavity field $a_{\bf k}$ to the input bath operators $F_{{\rm cav},12}$ and
$F^\dagger_{{\rm cav},12}$:
\begin{multline}
\left \langle a^{\dagger}_{\bf k}(t)  a_{\bf k}(t) \right \rangle =
\frac{1}{(2\pi)^2} \int_{-\infty}^{\infty} d\omega
\int_{-\infty}^{\infty} d\omega' \,\left\langle\tilde{a}^{\dagger}_{\bf
k}(\omega)
\tilde{a}_{\bf k}(\omega')\right \rangle~e^{-i(\omega-\omega')t} = \\
=\frac{1}{\pi} \int_{0}^{\infty} d\omega  |\bar{\mathcal
G}_{13}({{\bf k},-\omega})|^2  \Re \{\tilde{\Gamma}_{{{\rm cav}},\bf
k}(\omega)\} + |\bar{\mathcal G}_{14}({{\bf k},-\omega})|^2  \Re
\{\tilde{\Gamma}_{{12},\bf k}(\omega)\}.
\label{N_G}
\end{multline}
As the input bath is here taken in the vacuum state
$F_{{\rm cav},12}\,|0\rangle=0$, only the terms involving
$F^\dagger_{{\rm cav},12}$ contribute to the result (\ref{N_G}).
The presence of the anti-resonant light-matter coupling $H_{anti}$ is
here crucial, as it is responsible for the non-vanishing value of the
matrix elements ${\mathcal G}_{13}$ and $\bar{\mathcal G}_{14}$
connecting creation and destruction operators.

Remarkably, the ground state shows finite {\em anomalous} correlations
  between modes with opposite wavevectors:
\begin{multline}
\left\langle a_{\bf k}(t)  a_{-\bf k}(t) \right\rangle =
\frac{1}{(2\pi)^2} \int_{-\infty}^{\infty} d\omega
\int_{-\infty}^{\infty} d\omega' ~\left\langle \tilde{a}^{\dagger}_{\bf
k}(\omega)
\tilde{a}_{\bf k}(\omega') \right\rangle \,e^{i(\omega+\omega')t} = \\
=\frac{1}{\pi} \int_{0}^{\infty} d\omega \, \bar{\mathcal
G}_{11}({{\bf k},\omega}) \bar{\mathcal
G}_{13}({{-\bf k},-\omega}) \Re \{\tilde{\Gamma}_{{{\rm cav}},\bf
k}(\omega)\} + \bar{\mathcal
G}_{22}({{\bf k},\omega}) \bar{\mathcal G}_{14}({{-\bf k},-\omega})  \Re
\{\tilde{\Gamma}_{{12},\bf k}(\omega)\}~,
\label{cavity_anom}
\end{multline}
which suggest the ground state to be a sort of squeezed vacuum.

The predictions (\ref{N_G}) and (\ref{cavity_anom}) for respectively
the number of photons (solid line) and the anomalous correlations
(dashed line) have been plotted
in the left panel of Fig. \ref{fig_groundstate} as a function of the
normalized vacuum Rabi frequency $\Omega_R/\omega_{12}$ for the
resonant case $\omega_{{\rm cav},k} = \omega_{12}$.
For simplicity, in Fig.\ref{fig_groundstate} as well as in all the
following figures, a frequency-independent damping
of the cavity and electronic excitation modes has been used, i.e.,
$\Re\{\Gamma_{{\rm cav},\kk}(\omega > 0)\} = \bar{\Gamma}_{{\rm cav},k}$ and
$\Re\{\Gamma_{12,\kk}(\omega > 0)\} = \bar{\Gamma}_{12,k}$.
Note, however, that our theory is able to describe the system under
consideration for colored baths with an arbitrary frequency dependence
of $\Gamma_{{\rm cav},\kk}(\omega)$ and $\Gamma_{12,\kk}(\omega)$.

It is interesting to check that in the limit ${\tilde
\Gamma}_{{\rm cav},{\bf k}},{\tilde \Gamma}_{12,{\bf
k}}\rightarrow 0$ the values of (\ref{N_G}) and
(\ref{cavity_anom}) are in quantitative agreement with the result
of a direct diagonalization of the isolated cavity Hamiltonian
$H_{sys}$ as done in Ref.\onlinecite{Ciuti_vacuum}. Up to moderate
values of the broadening values, the difference from the isolated
cavity preduction remains small, in particular the total number of
virtual photons is only slightly changed.

Insight on the structure of the ground state can be obtained by
looking at other observables which are most sensitive to two-mode
squeezing effects.
 In the right panel of Fig. \ref{fig_groundstate} we have shown the
 results for the photonic correlation $C_{{\bf k},{-\bf
 k}}$ (solid line) between the noise amplitudes of the ${\bf k}$ and
 $-{\bf k}$-modes.  As usual, the amplitude quadrature operator of the ${\bf
   k}$-mode is defined as $\hat{X}_{\bf k} = a_{\bf k} + a_{\bf k}^{\dagger}$.
In the ground state, this correlation is
 \begin{equation}
 C_{{\bf k},{- \bf k}} = \frac{\langle (\hat{X}_{\bf k}- \langle \hat{X}_{\bf k}\rangle ) (\hat{X}_{-\bf k}- \langle \hat{X}_{-\bf k}\rangle )
 \rangle}{ \sqrt{\langle (\hat{X}_{\bf k}- \langle \hat{X}_{\bf k}\rangle )^2 \rangle \langle(\hat{X}_{-\bf k}- \langle \hat{X}_{-\bf k}\rangle )^2 \rangle}} = \frac{2 \Re \{  \langle a_{\bf k}  a_{-\bf k}
 \rangle\}}{1 + 2 \langle a^{\dagger}_{\bf k}  a_{\bf
 k}\rangle}~,
 \end{equation}
 where we have used the fact that in the ground state $\langle \hat{X}_{\bf k} \rangle = \langle \hat{X}_{-\bf k} \rangle = 0$.
As shown in right panel of Fig. \ref{fig_groundstate}, the
correlation tends to 0 when $\Omega_R/\omega_{12} \to 0$, i.e.,
the normal vacuum is reobtained in the weak-coupling limit.

In  the right panel of Fig. \ref{fig_groundstate}, the
normalized variance of the difference between
the amplitude quadratures of two correlated modes is shown, namely
\begin{equation}
F_{{\rm diff}} = \frac{1}{2}~\langle (\hat{X}_{\bf k}-\hat{X}_{-\bf k})^2 \rangle~ = 1 +  2 \langle a^{\dagger}_{\bf k}  a_{\bf
 k}\rangle - 2 \Re  \{  \langle a_{\bf k}  a_{-\bf k} \rangle \}~.
\end{equation}
Fluctuations below the shot noise level correspond to $F_{{\rm diff}} < 1$.
As shown in the figure, $F_{{\rm diff}}$ tends to $1$ for vanishing
$\Omega_R/\omega_{12}$.
When entering the ultra-strong coupling regime, $F_{{\rm diff}}$ decreases,
becoming smaller than 1 and the two-mode squeezing of the quantum
ground state becomes more and more significant.

It is important to stress that the
discussion in the present section involves the state of the intra-cavity field,
and that the virtual photonic and electronic excitations present in the
cavity ground state are trapped inside the cavity.
For an input field in the vacuum state, the output is in fact always
in the vacuum state as well and no radiation will be emitted, in
agreement with energy consevation requirements.
On the other hand, if the properties of the cavity system were modulated in time in a
non-adiabatic way, a creation of real excitations is possible
inside the cavity, which in turn leads to a finite emission of
{\em quantum vacuum radiation}\cite{Ciuti_vacuum} analogous to
what is predicted to happen in the so-called dynamical Casimir
effect\cite{dyn_Casimir,Lambrecht_review}. A complete study of these issues,
and in particular of the emission intensity, is presently the
subject of detailed investigations\cite{Deliberato}.

\section{Extracavity two-mode squeezing from a non-squeezed optical input ?}
\label{visibility}

An interesting question which has been raised in the literature
of interband excitonic
transitions~\cite{artoni,Schwendimann_bulk,Hradil} is whether any
quantum optical squeezing\cite{Walls} can be observed as a consequence
of the finite anomalous correlation shown by the polariton
vacuum~\cite{Hopfield}.
In our specific case, the question we ask is whether the finite anomalous
correlations (\ref{cavity_anom}) shown by the intracavity photon field
can be observed as a squeezing of the output field emitted by the
cavity. This, obviously,  in the absence of any squeezing in the input field.

The most general non-squeezed optical input state is identified by
the condition:
\begin{equation}
\left\langle \alpha_{q,\kk}^{in}\,\alpha_{q',\kk'}^{in}
  \right\rangle-
\left\langle \alpha_{q,\kk}^{in}\right\rangle\,\left\langle\alpha_{q',\kk'}^{in}
  \right\rangle=0,
\label{non_sq}
\end{equation}
which is well satisfied by a thermal incident state, as well as by
a coherent incident field\cite{Walls}.
The electronic input is assumed to be in a thermal state and to have
no correlations with the optical input.

As Eq. (\ref{unitary}) relates the annihilation operators of the
output to the annihilation operators of the input without
involving the creation ones, one has:
\begin{eqnarray}
\left\langle \alpha_{q,\kk}^{out}\right\rangle&=& \bar{\mathcal
  U}_{11}({{\bf k},\omega=\omega^{ph}_{q,\kk}})\,\left\langle
\alpha_{q,\kk}^{in}\right\rangle \\
\left\langle \alpha^{out}_{q,\kk}\,\alpha^{out}_{q',\kk'}\right\rangle&=& \bar{\mathcal
  U}_{11}({{\bf k},\omega=\omega^{ph}_{q,\kk}})\,\bar{\mathcal
  U}_{11}({{\bf k},\omega=\omega^{ph}_{q',\kk'}})\,\left\langle
\alpha_{q,\kk}^{in}\,\alpha_{q',\kk'}^{in}\right\rangle~,
\end{eqnarray}
from which it is immediate to prove that no squeezing is present in
the the output field either:
\begin{equation}
\left\langle \alpha_{q,\kk}^{out}\,\alpha_{q',\kk'}^{out} \right
\rangle- \left\langle
\alpha_{q,\kk}^{out}\right\rangle\,\left\langle\alpha_{q',\kk'}^{out}
  \right\rangle=0.
\end{equation}
This answers our initial question: no squeezing can be ever observed in
the output unless the input field is itself squeezed, or the
  properties of the cavity are modulated in time.

\section{Optical spectra}
\label{optical_spectra}
So far, the discussion has been limited to the case of a single
photonic bath coupled to the cavity. This model is sufficient to
describe a single-sided cavity in which the cavity mode is coupled to
the radiative modes through only one of its mirrors, while the other
one is supposed to be perfectly reflecting.
In order to obtain quantitative predictions for quantities of actual
experimental interest like reflection, transmission and absorption
spectra, one has to generalize the model adding a second photonic bath
so as to include the radiation emitted by the cavity through its back
mirror, as sketched in Fig.\ref{sketch}.

\subsection{Double-sided cavity: general theory}
\label{double_section}

The total Hamiltonian must now include a second photonic bath
associated to the radiation in the half-space behind the cavity, which
is coupled to the cavity mode through the back mirror.
This can be done by simply including another term analogous to (\ref{ph_bath}):
\begin{equation}
\label{ph_bath_pr} H^{ph'}_{bath} = \int dq~ \sum_{\bf k} \hbar
\omega^{ph}_{{q},{\bf k}} \left( \alpha'^{\dagger}_{{q},\bf k}
\alpha'_{{q},\bf k} + \frac{1}{2}\right) + i \hbar \int dq~
\sum_{\bf k} \left[\kappa^{ph'}_{{q},{\bf k}}\, \alpha'_{{q},{\bf
k}} \,a^{\dagger}_{\bf k}-\kappa^{ph'\,*}_{{q},{\bf
    k}}\,\alpha'^{\dagger}_{{q},{\bf k}}
\, a_{\bf k}\right].
\end{equation}
The new Hopfield matrix $\bar{\mathcal M}^{double}_{{\bf k},\omega}$
is obtained by $\bar{\mathcal M}_{{\bf k},\omega}$ in
(\ref{Hopfield_omega}) by simply replacing:
\begin{equation}
\tilde{\Gamma}_{{\rm cav},\bf k}(\omega) \to \tilde{\Gamma}^{T}_{{\rm cav},\bf k}(\omega)
= \tilde{\Gamma}_{{\rm cav},\bf k}(\omega) + \tilde{\Gamma}'_{{\rm cav},\bf k}(\omega)~,
\end{equation}
where $\tilde{\Gamma}'_{{\rm cav},\bf k}(\omega)~$ is the complex linewidth of the cavity
mode due to the finite transmittivity of the second mirror, defined in
a way analogous to the definition (\ref{Gamma_om}) of $\tilde{\Gamma}_{{\rm cav},\bf k}(\omega)$.
Correspondingly, a new Langevin force $F'_{{\rm cav},\bf k}(t)$ analogous to
(\ref{F_cav}) has to be included, which means to replace in (\ref{dyn_eq2})
\begin{equation}
\tilde{F}_{{\rm cav},\bf k}(\omega) \to \tilde{F}^{T}_{{\rm cav},\bf k}(\omega)
= \tilde{F}_{{\rm cav},\bf k}(\omega) + \tilde{F}'_{{\rm cav},\bf k}(\omega)~.
\end{equation}
Following the same steps as in the case of the one-sided cavity, the solutions for the output
operators are
\begin{equation}
  \left (
\begin{array}{c}
\alpha^{out}_{{q},{\bf k}} \\
\beta^{out}_{{q}',{\bf k}} \\
\alpha'^{out}_{{q},{\bf k}}
\end{array}
\right) =
{\mathcal U}^{double}({\bf k},\omega)
   \left (
\begin{array}{c}
\alpha^{in}_{{q},{\bf k}} \\
\beta^{in}_{{q}',{\bf k}} \\
\alpha'^{in}_{{q},{\bf k}}
\end{array}
\right)  ~, \label{unitary_double}
\end{equation}
where ${\mathcal U}^{double}({\bf k},\omega)$ is a $3 \times 3$ unitary
matrix.
Similarly to the single-sided case of (\ref{U11}-\ref{U22}), its
matrix elements can be written as:
\begin{equation}
\bar{\mathcal U}^{double}_{jl}({{\bf k},\omega}) =  \delta_{jl}
-  2 \Re
\{\tilde{\Gamma}_{l,\bf k}(\omega)\}
\left(\frac{\kappa^{j}_{{q},{\bf k}}}{\kappa^{l}_{{q}',{\bf
k}}}\right)^*\, \bar{\mathcal G}^{double}_{jl}({{\bf k},\omega}) ~.
\end{equation}
Here, a shorthand notation has been used: for $j=\{1,2,3\}$, the
  quantities $\tilde{\Gamma}_{j,\bf  k}(\omega)$ respectively mean
  $\tilde{\Gamma}_{{\rm cav},\bf   k}(\omega)$, $\tilde{\Gamma}_{12,\bf k}(\omega)$, and
  $\tilde{\Gamma}'_{{\rm cav},\bf  k}(\omega)$.
Analogously, $\kappa^{j}_{{q},{\bf k}}$ respectively mean
${\kappa^{ph}_{{q},{\bf k}}}$, ${\kappa^{el}_{{q},{\bf k}}}$, and
  ${\kappa'^{ph}_{{q},{\bf k}}}$ and all the three are evaluated at
  wavevector values such that $\omega^{ph}_{q,\kk}=\omega^{el}_{q',\kk}=\omega$.
$\bar{\mathcal G}^{double}({{\bf k},\omega})$ is defined as
  $\bar{\mathcal G}^{double}({{\bf k},\omega})= -i~[\bar{\mathcal
  M}^{double}_{{\bf k},\omega}]^{-1}~$.

In the single-sided cavity limit, one has $\bar{\mathcal
  U}^{double}_{jk}({{\bf k},\omega})=\bar{\mathcal U}_{jk}({{\bf
  k},\omega})$ and $\bar{\mathcal
  U}^{double}_{j3}({{\bf k},\omega})=
\bar{\mathcal U}^{double}_{3j}({{\bf k},\omega})=0$ for $j,k=\{1,2\}$,
  and $\bar{\mathcal
  U}^{double}_{33}({{\bf k},\omega})=1$.

\subsection{Linear optical spectra}
\label{reflection}
The results of the previous subsection can be used to obtain
quantitative predictions for the optical properties of the cavity,
namely its reflection, absorption and transmission spectra as a
function of the frequency $\omega$ of the incident light.
Here, we assume that no input other than optical is present, i.e.,
$\langle \beta^{in,\dagger}_{{q},{\bf k}} ~\beta^{in}_{{q},{\bf k}}
\rangle = 0$, and that the coherent radiation incides onto the cavity from the
half-space in front of it, i.e. $\langle
\alpha^{in,\dagger}_{{q},{\bf k}} ~\alpha^{in}_{{q},{\bf k}} \rangle >
0$, while $\langle \alpha'^{in,\dagger}_{{q},{\bf k}}
~\alpha'^{in}_{{q},{\bf k}} \rangle = 0$.

In the considered geometry, reflection is described by the output operator
$\alpha^{out}_{{q},{\bf k}}$, so that the reflectivity is equal to
\begin{equation}
{\mathcal R}_{\bf k}(\omega) =  \frac{\langle \alpha^{out
\dagger}_{{q},{\bf k}} ~\alpha^{out}_{{q},{\bf k}}
\rangle}{\langle \alpha^{in \dagger}_{{q},{\bf k}}
~\alpha^{in}_{{q},{\bf k}} \rangle} =|\bar{\mathcal
U}^{double}_{11}({{\bf k},\omega})|^2~.
\end{equation}
The transmission through the second mirror is described by the
output operator $\alpha'^{out}_{{q},{\bf k}}$. Therefore, the
transmittivity reads
\begin{equation}
{\mathcal T}_{\bf k}(\omega) = \frac{\langle \alpha'^{out
\dagger}_{{q},{\bf k}} ~\alpha'^{out}_{{q},{\bf k}}
\rangle}{\langle \alpha^{in \dagger}_{{q},{\bf k}}
~\alpha^{in}_{{q},{\bf k}} \rangle} =  |\bar{\mathcal
U}^{double}_{31}({{\bf k},\omega})|^2~.
\end{equation}
Note that ${\mathcal T}_\kk(\omega)=0$ in the single-sided case.
The presence of the electronic bath implies that the incident
light can be absorbed into electronic energy. The corresponding
absorption coefficient of the microcavity system is:
\begin{equation}
{\mathcal A}_{\bf k}(\omega) =  |\bar{\mathcal
U}^{double}_{21}({{\bf k},\omega})|^2~.
\end{equation}
As expected, the total energy is conserved, i.e. ${\mathcal
R}_{\bf k}(\omega) + {\mathcal A}_{\bf k}(\omega) + {\mathcal
T}_{\bf k}(\omega) = 1$, as one can verify from the unitarity of
$\bar{\mathcal U}^{double}({{\bf k},\omega})$.

\begin{figure}[t!]
\begin{center}
\includegraphics[width= 9 cm]{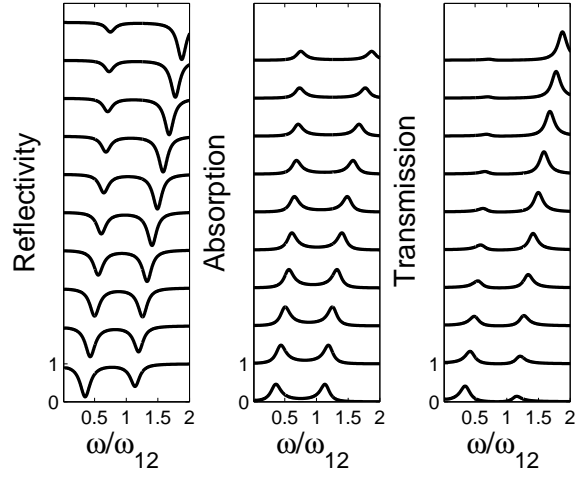}
 \caption{\label{fig_reflectivity} Reflectivity ${\mathcal
R}_{\bf k}(\omega)$ (left), absorption ${\mathcal A}_{\bf k}(\omega)$ (center) and
transmission ${\mathcal T}_{\bf k}(\omega)$ (right)
    spectra as a
   function of the normalized photon
frequency $\omega/\omega_{12}$ for different values of the
normalized detuning $\delta = (\omega_{{\rm cav},k} -
\omega_{12})/\omega_{12}$ (from $\delta = -0.5$ to $\delta = +0.5$
by steps of $0.1$). The different curves are offset for clarity.
The bottom curve corresponds to $\delta = - 0.5$, while the top
one has been obtained for $\delta = 0.5$. Parameters:
$\Omega_{R,k}/\omega_{12} = 0.4$. Broadening parameters as in Fig.
\ref{fig_groundstate}}.
\end{center}
\end{figure}

\begin{figure}[t!]
\begin{center}
\includegraphics[width= 9 cm]{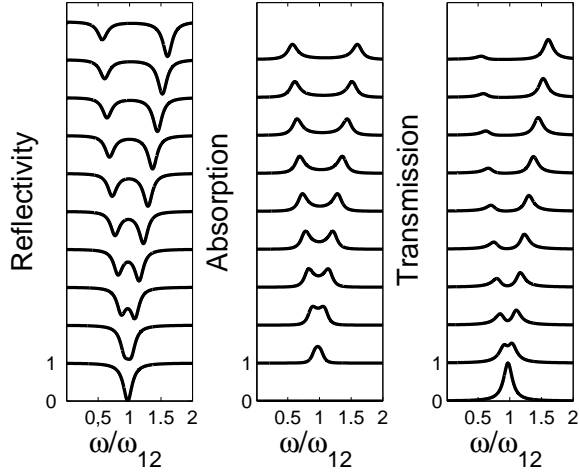}
\caption{ \label{fig_splitting}
Reflectivity ${\mathcal
R}_{\bf k}(\omega)$ (left), absorption ${\mathcal A}_{\bf k}(\omega)$ (center)
and transmittivity ${\mathcal T}_{\bf
  k}(\omega)$ (right)
    spectra as a function of
the normalized photon frequency $\omega/\omega_{12}$ for different
values of the normalized vacuum Rabi frequency
$\Omega_R/\omega_{12}$ for the resonant case $\omega_{{\rm cav},k} =
\omega_{12}$. The different curves are offset for clarity. The
bottom curve corresponds to $\Omega_R = 0$, while the top curve
corresponds to $\Omega_R = 0.5\omega_{12}$
($\Omega_R/\omega_{12}$ is increased by steps of $0.05$). Same
broadening parameters as in Fig.\ref{fig_reflectivity}.}
\end{center}
\end{figure}

In Fig. \ref{fig_reflectivity}, we show an example of reflectivity,
transmission, and absorption spectra for different values of the normalized
detuning $\delta = (\omega_{{\rm cav},k}-\omega_{12})/\omega_{12}$,
which is varied from $-0.5$ to $0.5$ by steps equal to 0.1. The
linear optical spectra have resonances corresponding to the cavity
polariton eigenmodes.
For the large vacuum Rabi frequency here
considered ($\Omega_R = 0.4 \omega_{12}$), it is apparent that the
optical spectra show an anticrossing of the polariton eigenmodes,
which is is strongly asymmetric, as it was anticipated in the
case of a closed cavity system \cite{Ciuti_vacuum} without dissipation.
This can be also seen in Fig. \ref{fig_splitting}, showing the
reflection, transmission, and absorption spectra for zero detuning
($\omega_{{\rm cav},k}= \omega_{12}$) and increasing values of
$\Omega_R/\omega_{12}$ (from $0$ to $0.5$ with steps of $0.05$).
Remarkably, note that for an incident frequency close to resonance
with the polariton eigenmodes, a significative fraction of the
incident energy goes is dissipated in the electronic bath as
absorption.


\subsection{Electroluminescence}
\label{electroluminescence}
\begin{figure}[t!]
\begin{center}
\includegraphics[width= 8 cm]{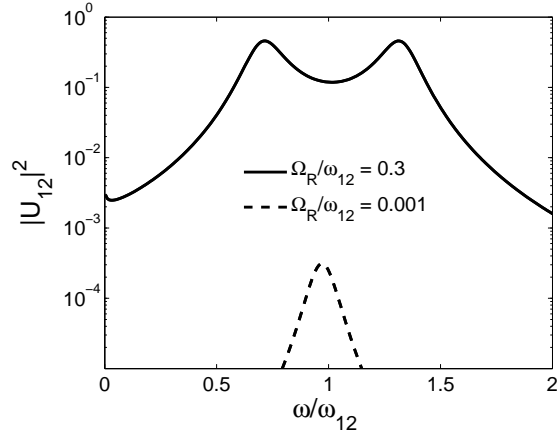}
\caption{ \label{fig_U_12} A logarithmic scale plot of the
dimensionless quantity $|\bar {\mathcal U}^{double}_{12}(\kk,\omega)|^2$
(proportional to the electroluminescence spectrum) as a function
of the normalized frequency $\omega/\omega_{12}$ in a
  weak-coupling regime $\Omega_R/\omega_{12} = 0.001$ (dashed-line)
and  in a strong-coupling one $\Omega_R/\omega_{12} = 0.3$
(solid line). Other parameters as in
Fig. \ref{fig_splitting}.
The small increase at very low $\omega$ is a consequence of the
  specific choice of constant values for $\Re \{\tilde{\Gamma}_{{{\rm cav}},\bf k}(\omega > 0) \}$ and
  $\Re \{\tilde{\Gamma}_{{12},\bf k}(\omega > 0) \}$, producing a Kramers-Kronig
  singularity of the imaginary part (Lamb shift) at $\omega =0$.
}
\end{center}
\end{figure}

\begin{figure}[t!]
\begin{center}
\includegraphics[width= 8 cm]{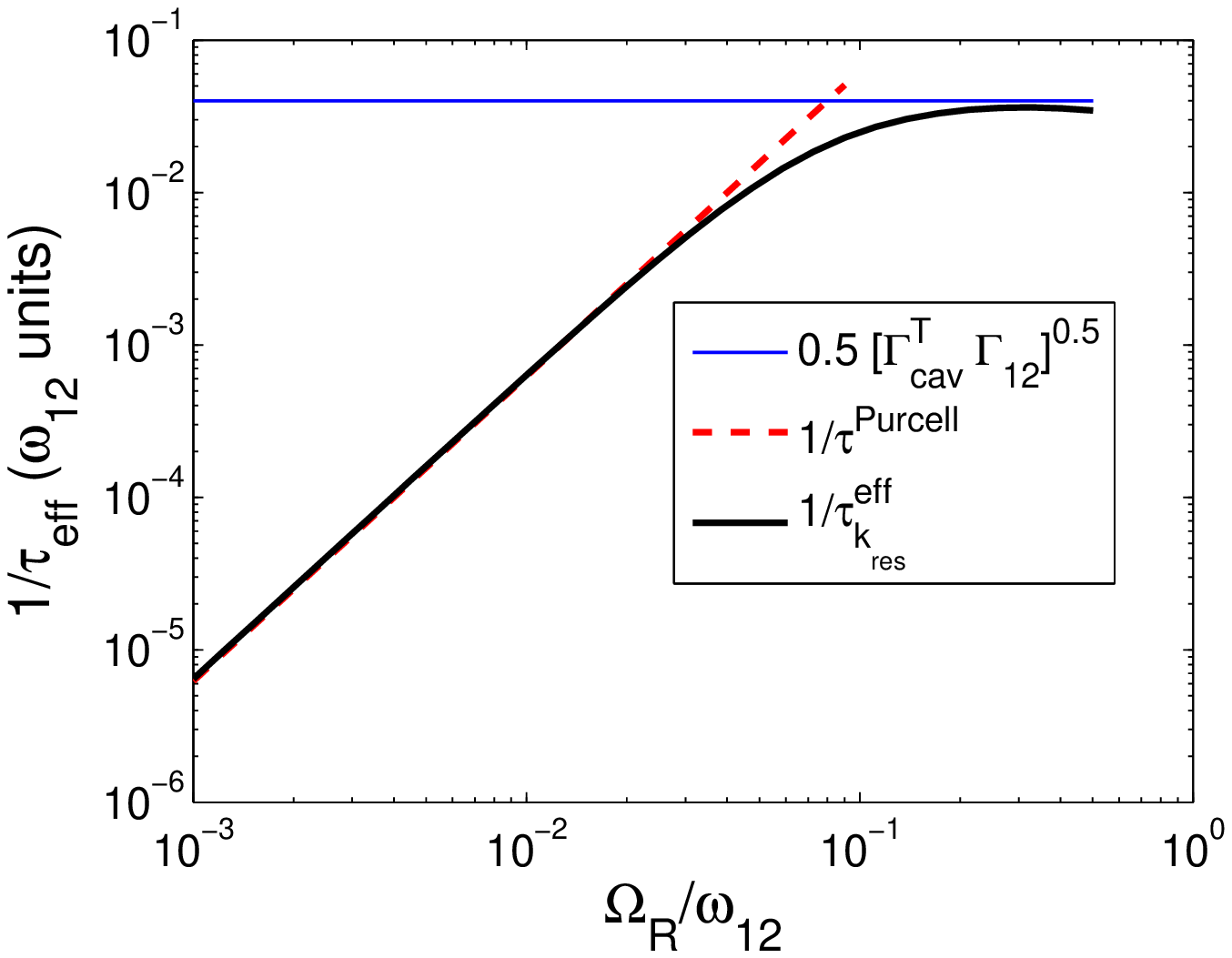}
\caption{ \label{fig_purcell} Solid line: the effective spontaneous emission rate
$\gamma^{{\rm eff}}_k$ (in units of $\omega_{12}$) as a function of the normalized
vacuum Rabi frequency $\Omega_{R,k}/\omega_{12}$ for $\omega_{{\rm
    cav},k} = \omega_{12}$. Dashed-line: the Purcell rate
$2 \frac{\bar \Gamma_{{\rm cav}}}{\bar \Gamma_{{\rm
      cav}}^T} \frac{\Omega_{R,k}^2}{\bar \Gamma_{{\rm cav}}^T + \bar
  \Gamma_{12}}$. Horizontal line: the strong coupling limit
$\frac{1}{2}\,\sqrt{\Gamma_{{\rm cav}}^T \Gamma_{12}}$. Parameters:
${\bar \Gamma}_{{\rm cav},\kk} = {\bar \Gamma}'_{{\rm cav},\kk} = 0.04
\omega_{12}$ and ${\bar \Gamma}_{12,\kk} = 0.08 \omega_{12}$.}
\end{center}
\end{figure}
Another physical quantity which can be successfully studied by the
present theory is the intensity of the light emitted by the cavity
when this is incoherently excited from the electronic bath in a
  so-called {\em electroluminescence} experiment.
Electrically excited intersubband transitions are in fact playing an
important role as light sources in the mid and in the far infrared
region, but they still suffer from a poor quantum efficiency of
radiative emission as compared to non-radiative losses.
A way of enhancing the emission efficiency would therefore be of great
interest.

Consider a purely electronic, incoherent input,
  distributed among the different $q$ modes:
\begin{eqnarray}
\langle \beta^{in,\dagger}_{{q},{\bf k}} ~\beta^{in}_{{q},{\bf k}}
  \rangle & =& I^{el}_{exc,q}> 0 \label{eq:electr_excit} ~,\\
\langle \alpha'^{in,\dagger}_{{q},{\bf k}} \, \alpha'^{in}_{{q},{\bf
  k}} \rangle &=&  \langle \alpha'^{in,\dagger}_{{q},{\bf k}}
  \, \alpha'^{in}_{{q},{\bf k}} \rangle = 0.
\end{eqnarray}
The intensity of the spontaneously emitted light through the front
mirror (defined as the mean number of photons in the $q$ mode) is
proportional to the quantity
\begin{equation}
{\mathcal L}_{\kk}= ~\int\! dq ~\langle \alpha^{out,
\dagger}_{{q},{\bf k}} ~\alpha^{out}_{{q},{\bf k}} \rangle =
 \int \! d\omega~ \rho^{ph}_{\kk}(\omega)~ |\bar{\mathcal
U}^{double}_{12}({{\bf k},\omega})|^2  I^{el}_{exc}(\omega) .
\label{emission_complete}
\end{equation}
The dimensionless quantity $|\bar{\mathcal U}^{double}_{12}({{\bf
k},\omega})|^2$ is represented in Fig. \ref{fig_U_12} for a
weak-coupling situation (dashed-line) and for a strong coupling
case (solid line), where it exhibits two polaritonic resonances.
 Suppose that in the spectral region where $|\bar{\mathcal
U}^{double}_{12}({{\bf k},\omega})|^2$ is significant we can
roughly neglect the frequency-dependence of the electronic excitation
($I^{el}_{exc}(\omega) \approx I^{el}_{exc}$) and
of the extracavity photon density of state ($\rho^{ph}_{\kk}(\omega)
\approx \rho^{ph}_{\kk})$.
With these approximations, the luminescence intensity is given by the simplified expression
\begin{equation}
{\mathcal L}_{\kk} \approx \rho^{ph}_{\kk} I^{el}_{exc}
{\gamma^{{\rm eff}}_{\kk}}~,
\end{equation}
where the effective rate of the luminescence is
\begin{equation}
{\gamma^{{\rm eff}}_{\kk}} = \int \!d\omega~ |\bar{\mathcal
 U}^{double}_{12}({{\bf k},\omega})|^2  .
\end{equation}

If we further neglect the frequency-dependence of the broadening,
(i.e. $\Re\{\Gamma_{{\rm cav},\kk}(\omega > 0)\} \approx \bar{\Gamma}^T_{{\rm cav},\kk}$,
$\Re\{\Gamma_{12,\kk}(\omega > 0)\} \approx \bar{\Gamma}_{12,\kk}$)
and consider the resonant case $\omega_{{\rm cav},k} = \omega_{12}$, the results
can be expressed by simple analytical expressions.
In the weak-coupling regime $\Omega_{R,k}^2\ll \bar{
    \Gamma}^{T}_{{\rm cav},\kk} \bar{\Gamma}_{12,\kk}$, it can be
accurately approximated by a Purcell-like\cite{Purcell,Gerard} law:
\begin{equation}
{\gamma^{{\rm eff}}_{\kk}} \simeq  \, 2\, \frac{\bar{
\Gamma}_{{\rm cav},\kk}}{{\bar \Gamma}^T_{{\rm cav},\kk}}~
\frac{\Omega_{R,\kk}^2}{{\bar \Gamma}^T_{{\rm cav},\kk}+ \bar \Gamma_{12,\kk}},
\label{weakemiss}~
\end{equation}
where we recall that $\bar \Gamma^T_{{\rm cav},\kk} = \bar \Gamma_{{\rm cav},\kk}+ \bar \Gamma'_{{\rm cav},\kk} $ is
the total cavity broadening (due to the front and back mirror).
In contrast, in the ultra-strong coupling regime, ${\gamma^{{\rm eff}}_{\kk}}$
saturates around the value
\begin{equation}
{\gamma^{{\rm eff}}_{\kk}} \simeq  \frac{1}{2} \sqrt{{\bar
\Gamma}_{{\rm cav},\kk}^T {\bar \Gamma}_{12,\kk}} \label{ultrastrongemiss}
\end{equation}
The numerical dependence of ${\gamma^{{\rm eff}}_{\kk}}$ as a
function of $\Omega_{R,\kk}/\omega_{12}$ is shown in Fig. \ref{fig_purcell}, showing the
Purcell-like behavior (\ref{weakemiss}) in the weak-coupling limit and the saturation
value (\ref{ultrastrongemiss}) in the very strong coupling limit.
The slight decrease at large $\Omega_{R,\kk}$ is due to the specific
shape chosen for the real parts of the damping kernels $\Gamma_{{\rm cav},\kk}(\omega)$ and
$\Gamma_{12,\kk}(\omega)$, which are constant for
$\omega>0$ and vanishing for $\omega<0$, with a jump at $\omega=0$.
Due to the Kramers-Kronig relationship, the imaginary part (the Lamb shift) has a
singularity at $\omega = 0$.

\subsection{Comparison with free space case: enhancement of electroluminescence}
\label{sec:applications}

It is interesting to compare the predictions for the ultra-strong
coupling regime to what occurs for the isolated quantum well in
the absence of the surrounding microcavity. In order for the
comparison of the electroluminescence rates to
  be fair, exactly the same model (\ref{eq:electr_excit}) has to
be used in both cases for the excitation of the intersubband
transition by the electronic bath. A huge Hilbert space is in fact
available for the electronic excitations of the present
many-electron system, and the spontaneous
  emission rate dramatically depends on the specific state under
  consideration
In our model, the only bright states are the ones created by the
action of the electronic polarization field creation operators
$b^\dagger_{\bf k}$, while the much larger number of other states
remain dark. This is in stark contrast with what happens in a
two-level atom, where the presence of a single excited state makes
the spontaneous emission rate to be an univocally defined
quantity. In particular, an explicit calculation of the quantum
efficiency of the electroluminescence process with a realistic
model of the electronic injection process would require a more
refined model taking into account the energy that flows into all
the dark excited states, orthogonal to the bright mode. For this
reason, we do not attempt to estimate it in the present paper, but
we rather focus our attention on comparing the predictions for the
emission intensities that are obtained in the two cases using the
very same model for the electronic injection mechanism.

In the absence of the surrounding microcavity, no coupling to any
cavity mode is present, and the intersubband transition is
directly coupled to free-space radiative modes.
For the frequency-flat excitation of the electronic bath
(\ref{eq:electr_excit}), the emission intensity can be analytically
calculated to be:
\begin{equation}
{\mathcal L}^{QW}_{\kk} =  \rho^{ph}_{\kk}\, I^{el}_{exc}\,
{\gamma_{{\rm bright},\kk}^{QW}}~,
\end{equation}
where $\gamma_{{\rm bright},\kk}^{QW}$ is spontaneous emission rate of the
bright intersubband excitation of wavevector $\kk$ when the quantum well is
embedded in a bulk
material of refractive index $\epsilon_{QW}$ without any surrounding
cavity.

For an electronic surface density in the quantum well equal to
$\sigma_{el}$, the free-space spontaneous emission rate of the bright
excitation mode is calculated by applying the Fermi's golden rule, giving
the result
\begin{equation}
\gamma_{{\rm bright},\kk}^{QW}=
\frac{1}{2 \sqrt{\epsilon_{QW}}}\,\frac{e^2}{c}\,\frac{N_{QW} \sigma_{el}}{2m_0}\,f_{12}\,\frac{\sin^2\theta}{\cos\theta}
\end{equation}
where $e$ is the electron charge, $m_0$ is the free electron mass, $c$
the speed of light and $N_{QW}$ the number of identical quantum wells.
The propagation angle $\theta$ of the emitted photon is given by $\sin\theta=c k /(\omega_{12}
\sqrt{\epsilon_{QW}})$.
The oscillator strength $f_{12}$ of the intersubband transition
is written in terms of the electric dipole matrix element $z_{12}$
as~\cite{Sirtori94}:
\begin{equation}
f_{12}=\frac{2m_0\,\omega_{12}\, z_{12}^2}{\hbar}.
\end{equation}
Note that the bright state $b^\dagger_{\bf k}\,|F\rangle$ is the
excited state with the largest spontaneous emission rate. In
particular, $\gamma_{{\rm bright},\kk}^{QW}$ increases with the
density of the two-dimensional electron gas and does not depend on
the emission frequency $\omega_{12}$. It is interesting to compare
$\gamma_{{\rm bright},\kk}^{QW}$ with the spontaneous emission
rate of a different excited state of the two-dimensional electron
gas, namely $c^{(j) \dagger}_{2, \kk + {\bf q}} c^{(j)}_{1, \kk }
|F\rangle$. In this case, the intersubband excitation does not
coherently involve any longer all the electrons in the lower
subband, but only the one initially at $\kk$, so that the
corresponding spontaneous emission rate is suppressed by a factor
equal to the total number $N_{QW}N_{el}$ of electrons. Note that
an atomic-like spontaneous emission rate\cite{Rosencher} (i.e.,
proportional to $z_{12}^2 \omega_{12}^3$ and very weak for long
wavelength transitions in the far infrared) is achieved only for a
state like $c^{\dagger (j)}_{2, \kk} |0_{cond}\rangle$ (one
electron in the second subband, none in the first one). In
contrast, the strong radiative properties of the bright state
$b^\dagger_{\bf k} |F>$ with spontaneous emission rate
$\gamma_{{\rm bright},\kk}^{rad}$ occur in presence of a dense
electron gas, as considered in this paper.

For a quantum well with very high barriers $f_{12}$ does not
depend on the quantum well thickness (i.e. on $\omega_{12}$) and
is approximately given by $f_{12}\simeq 0.96\,m_0/m^*$, $m^*$
being the effective electronic mass in the
semiconductor~\cite{bastard}. Using GaAs parameters
($\epsilon_{QW} = 13.5$, $f_{12}\simeq 14$), with an electron
density $\sigma_{el}=5\cdot 10^{11}\,\textrm{cm}^{-2}$ and $N_{QW}
= 10$, the radiative linewith of the bright state is approximately
$\hbar\gamma^{QW}_{{\rm bright},\kk} \simeq 0.04\,\textrm{meV}$.

Comparing this result with the one (\ref{ultrastrongemiss}) for the ultra-strong
coupling regime, the cavity-induced enhancement $\eta_{\kk}$ of
the spontaneous emission of the quantum well is given at resonance
$\omega_{\kk}=\omega_{12}$ by
\begin{equation}
\label{enhancement} \eta_{\kk} \equiv \frac{{\mathcal
L}_{\kk}}{{\mathcal L}^{QW}_{\kk}} = \frac{\gamma^{{\rm
eff}}_{\kk}} {\gamma_{{\rm bright},\kk}^{QW}} \approx
\frac{\sqrt{{\bar \Gamma}^T_{{\rm cav},\kk} {\bar
\Gamma}_{12,\kk}}}{2\, \gamma_{{\rm bright},\kk}^{rad}} ~.
\end{equation}
For an intersubband transition for which $\hbar \omega_{12} \approx
100$ meV, typical values for the non-radiative broadening and the
cavity mode broadening are $\hbar \Gamma_{12,\kk},\hbar
\Gamma_{{\rm cav},\kk} \approx 10\,\textrm{meV}$.
The enhancement is then as large as $\eta_{\kk} \approx 100$.

The fact that (\ref{enhancement}) holds only in the strong coupling regime
$\sqrt{{\bar \Gamma}^T_{{\rm cav},\kk} {\bar \Gamma}_{12,\kk}} \ll \Omega_{R,\kk}$
imposes that:
\begin{equation}
\label{enhancement_boundary}
\eta_{\kk} \ll \frac{\Omega_{R,\kk}}{\gamma^{rad}_{{\rm bright},\kk}}~,
\end{equation}
which means that the enhancement can not made arbitrary large by
simply choosing larger linewidths.

As a final remark, note that the expression (\ref{enhancement}) has
been obtained under the resonant condition $\omega_{{\rm cav},\kk} =
\omega_{12}$ and is therefore expected to hold in a cone of
wavevectors $\kk$ around the resonant wavevector $k_{res}$ such as
$\omega_{{\rm cav},k_{res}} = \omega_{12}$.
On the other hand, when the cavity mode resonance is detuned from the
intersubband transition resonance of an amount larger than the
linewidths, the enhancement with respect to free space case
disappears.

\section{Conclusions and perspectives}
\label{conclusions}
In conclusion, we have presented a full quantum theory
of cavities in the ultra-strong light-matter coupling regime (i.e.
when the vacuum Rabi frequency is comparable to the
active electronic transition frequency), including the coupling
of the cavity system to dissipative baths.
In the case of a time-independent cavity properties,
we have solved exactly the quantum Langevin equations for the
intracavity operators within an input-output approach and we have
determined analytically the output operators, allowing us to determine
the response of the cavity system to an arbitrary input.

In the case of a vacuum input for the photonic and electronic
polarization fields, we have characterized the properties of the
ground state of the cavity system: due to the anti-resonant terms
of the light-matter interaction, it turns out to be a two-mode
squeezed vacuum, whose properties are weakly renormalized by the
interaction with the photonic and electronic baths. In particular,
we have checked that the photons and electronic excitations
present in the ground state are purely virtual ones, and can not
escape from the cavity: if the input is in the vacuum state, the
output is itself in the vacuum state and no radiation is emitted
by the system. Furthermore, we have explicitly shown that if no
anomalous correlations are present in the input beam, no
correlations will be present in the output either, so that no
extra-cavity squeezing is observable unless the input is itself
squeezed.

On the other hand, the anomalous properties due to the
antiresonant terms of the large vacuum Rabi coupling show up as a
peculiar asymmetric anticrossing of the polariton excitation
branches that can be easily observed in the optical reflection,
transmission, and absorption spectra under coherent light
excitation.

Finally, the input-output formalism has been applied to the study
of the electroluminescence emission intensity in the case of an
electronic excitation: the use of a microcavity surrounding the
quantum well provides a significant enhancement of the emission
performances as compared to the ones of an isolated quantum well.

As a future perspective, the theory presented here can be
generalized to explore the interesting and fascinating scenario of
a time-modulated cavity. Even in the case of a vacuum input for
the photonic and electronic polarization fields, a non-adiabatic
temporal variation of the vacuum Rabi frequency (which is possible
in the case of intersubband transitions in doped quantum
well\cite{Aji_APL}) is expected to produce an output radiation of
correlated photons\cite{Ciuti_vacuum}, an effect reminiscent of
the dynamical Casimir effect. The calculation of the quantum
vacuum radiation spectra in the modulated case will be possible
using and generalizing the comprehensive formalism developed in
this paper.

\acknowledgments We wish to thank A. Anappara, G. Bastard, V.
Berger, Y. Castin, R. Colombelli, S. De Liberato, C. Fabre, L.
Sapienza, C. Sirtori, A. Tredicucci, A. Vasanelli, A. Verger for
discussions. LPA-ENS is a "Unit\'{e} Mixte de Recherche
Associ\'{e} au CNRS (UMR 8551) et aux Universit\'{e}s Paris 6 et
7".

%%%%%%%%%%%%%%%%%%%%%%%%%%%%%%%%%%%%%%%%%%%%%%%%%%%%%%%%%%%%%%%%%%%%%
%%%%%%%%%%%%%%%%%%%%%%%%%%%%%%%%%%%%%%%%%%%%%%%%%%%%%%%%%%%%%%%%%%%%%
%%%%% APPENDIX %%%%%%%%%%%%%%%%%%%%%%%%%%%%%%%%%%%%%%%%%%%%%%%%%%%%%%
%%%%%%%%%%%%%%%%%%%%%%%%%%%%%%%%%%%%%%%%%%%%%%%%%%%%%%%%%%%%%%%%%%%%%
%%%%%%%%%%%%%%%%%%%%%%%%%%%%%%%%%%%%%%%%%%%%%%%%%%%%%%%%%%%%%%%%%%%%%
\appendix
\section{Generalization of the input-output theory including also the anti-resonant coupling
terms in the bath Hamiltonian} \label{General}

In this Appendix, we generalize the input-output theory
of the present system in order to include also the anti-resonant terms
in the coupling with the photonic and electronic baths.
These additional Hamiltonian
terms make the calculations more cumbersome, but the results are
still analytical. We have verified explicitly that when the
broadening of the cavity and electronic transition modes is
moderate, the effect of the anti-resonant terms in the bath Hamiltonians
are negligible. The bath Hamiltonians including the antiresonant
terms are
\begin{eqnarray}
H^{ph,gen}_{bath} =  H^{ph}_{bath}+ i \hbar \int dq \sum_{\bf k}
\left(\kappa^{ph}_{{q},{\bf k}}\,\alpha_{{q},{\bf k}} ~a_{\bf k}-
\kappa^{ph\,*}_{{q},{\bf k}}\,\alpha^{\dagger}_{{q},{\bf k}}
a^{\dagger}_{\bf k} \right)~,
\end{eqnarray}
and
\begin{eqnarray}
H^{el,gen}_{bath} =  H^{el}_{bath} + i \hbar \int dq \sum_{\bf k}
\left(\kappa^{el}_{{q},{\bf k}}\,\beta_{{q},{\bf k}} ~b_{\bf k}-
\kappa^{el\,*}_{{q},{\bf k}}\,\beta^{\dagger}_{{q},{\bf k}}
b^{\dagger}_{\bf k} \right)~,
\end{eqnarray}
The generalized damping kernels
\begin{equation}
 \gamma_{{\rm cav}, {\bf k}}(t) = \Gamma_{{\rm cav},{\bf k}}(t) -
\Gamma^{\dagger}_{{\rm cav},-{\bf k}}(t)~, \end{equation}
\begin{equation}
 \gamma_{12, {\bf k}}(t) = \Gamma_{12,{\bf k}}(t) -
\Gamma^{\dagger}_{12,-{\bf k}}(t)~, \end{equation}
and the generalized Langevin forces
\begin{equation}
f_{{\rm cav}, {\bf k}}(t) = F_{{\rm cav},{\bf k}}(t) - F^{\dagger}_{{\rm cav},-{\bf
k}}(t)~,
\end{equation}
\begin{equation}
f_{12, {\bf k}}(t) = F_{12,{\bf k}}(t) - F^{\dagger}_{12,-{\bf
k}}(t)~.
\end{equation}
The quantum Langevin equations in frequency space become
\onecolumngrid
\begin{equation}
 \bar{\mathcal M}^{gen}_{{\bf k},\omega} ~ \left (
\begin{array}{c}
\tilde{a}_{\bf k}(\omega) \\
\tilde{b}_{\bf k}(\omega) \\
\tilde{a}^{\dagger}_{-\bf k}(-\omega) \\
\tilde{b}^{\dagger}_{-\bf k}(-\omega) \\
\end{array}
\right) + i \left (
\begin{array}{c}
\tilde{f}_{{\rm cav},{\bf k}}(\omega) \\
\tilde{f}_{12,{\bf k}}(\omega) \\
- \tilde{f}_{{\rm cav},{\bf k}}(\omega) \\
- \tilde{f}_{12,{\bf k}}(\omega) \\
\end{array}
\right) = 0 ~,
\label{dyn_eq_compl}
\end{equation}
where
\begin{equation}
\bar{\mathcal M}^{gen}_{{\bf k},\omega} =  \left (
\begin{array}{cccc}
\omega_{{\rm {\rm cav}},k} + 2 {\bar{D}}_k-\omega - i \tilde{\gamma}_{{\rm cav},{\bf k}}(\omega)  & i \bar{\Omega}_{R,k} & 2{\bar{D}}_k - i \tilde{\gamma}_{{\rm cav},{\bf k}}(\omega)& -i \bar{\Omega}_{R,k} \\
-i\bar{\Omega}_{R,k} & \omega_{12} -\omega - i \tilde{\gamma}_{12,{\bf k}}(\omega) & - i\bar{\Omega}_{R,k} & - i \tilde{\gamma}_{12,{\bf k}}(\omega) \\
-2{\bar{D}}_k + i \tilde{\gamma}_{{\rm cav},{\bf k}}(\omega)& -i\bar{\Omega}_{R,k} & -\omega_{{\rm cav},k} - 2 {\bar{D}}_k -\omega + i \tilde{\gamma}_{{\rm cav},{\bf k}}(\omega) & i \bar{\Omega}_{R,k} \\
-i\bar{\Omega}_{R,k} &  i \tilde{\gamma}_{12,{\bf k}}(\omega) & -i\bar{\Omega}_{R,k} &  -\omega_{12} -\omega + i \tilde{\gamma}_{12,{\bf k}}(\omega)\\
\end{array}
\right )
\end{equation}
%\twcolumngrid The generalized input-out boundary conditions for
the output operators $\alpha^{out}_{{q},{\bf k}}$ and
$\beta^{out}_{{q},{\bf k}}$ are related to the input and cavity ones by:
\begin{equation}
\label{io_exp_1} \alpha^{out}_{{q},{\bf k}} =
\alpha^{in}_{{q},{\bf k}}  - \kappa^{ph\star}_{{q},{\bf k}}
(\tilde{a}_{\bf k}(\omega^{ph}_{{q},{\bf k}}) +
\tilde{a}^{\dagger}_{-\bf k}(-\omega^{ph}_{{q},{\bf k}}))~,
\end{equation}
\begin{equation}
\label{io_exp_2} \beta^{out}_{{q},{\bf k}} = \beta^{in}_{{q},{\bf
k}}  - \kappa^{el\star}_{{q},{\bf k}} (\tilde{b}_{\bf
k}(\omega^{el}_{{q},{\bf k}}) + \tilde{b}^{\dagger}_{-\bf
k}(-\omega^{el}_{{q},{\bf k}}))~.
\end{equation}

As previously, Eq. (\ref{dyn_eq_compl}) can be solved exactly:
\begin{equation}
  \left (
\begin{array}{c}
\tilde{a}_{\bf k}(\omega) \\
\tilde{b}_{\bf k}(\omega) \\
\tilde{a}^{\dagger}_{-\bf k}(-\omega) \\
\tilde{b}^{\dagger}_{-\bf k}(-\omega) \\
\end{array}
\right) =  \bar{\mathcal G}^{gen}({{\bf k},\omega})  \left (
\begin{array}{c}
\tilde{f}_{{\rm cav},{\bf k}}(\omega) \\
\tilde{f}_{12,{\bf k}}(\omega) \\
- \tilde{f}_{{\rm cav},{\bf k}}(\omega) \\
- \tilde{f}_{12,{\bf k}}(\omega) \\
\end{array}
\right)  ~,
\end{equation}
where
\begin{equation}
\bar{\mathcal G}^{gen}({{\bf k},\omega}) = -i~[\bar{\mathcal
M}^{gen}_{{\bf k},\omega}]^{-1}~.
\end{equation}

The relation between the input and output operators is
\begin{equation}
  \left (
\begin{array}{c}
\alpha^{out}_{{q},{\bf k}} \\
\beta^{out}_{{q}',{\bf k}} \\
\end{array}
\right) = \left (
\begin{array}{cc}
\bar{\mathcal U}^{gen}_{11}({{\bf k},\omega}) &  \bar{\mathcal U}^{gen}_{12}({{\bf k},\omega})\\
\bar{\mathcal U}^{gen}_{21}({{\bf k},\omega}) &  \bar{\mathcal U}^{gen}_{22}({{\bf k},\omega})\\
\end{array}
\right)
   \left (
\begin{array}{c}
\alpha^{in}_{{q},{\bf k}} \\
\beta^{in}_{{q}',{\bf k}} \\
\end{array}
\right)  ~,
\end{equation}
with the generalized matrix \onecolumngrid
\begin{eqnarray}
\bar{\mathcal U}^{gen}_{11}({{\bf k},\omega}) =  1-  2 \Re
\{\tilde{\Gamma}_{{{\rm cav}},\bf k}(\omega)\} \left [\bar{\mathcal
G}^{gen}_{11}({{\bf k},\omega})
          + \bar{\mathcal G}^{gen}_{31}({{\bf k},\omega})
          - \bar{\mathcal G}^{gen}_{13}({{\bf k},\omega})
          - \bar{\mathcal G}^{gen}_{33}({{\bf k},\omega})\right ] ~,\\
\bar{\mathcal U}^{gen}_{12}({{\bf k},\omega}) =  -  2 \Re
\{\tilde{\Gamma}_{{12},\bf k}(\omega)\}
\frac{\kappa^{ph~\star}_{{q},{\bf
k}}}{\kappa^{el~\star}_{{q}',{\bf k}}}  \left [\bar{\mathcal
G}^{gen}_{12}({{\bf k},\omega})
          + \bar{\mathcal G}^{gen}_{32}({{\bf k},\omega})
          - \bar{\mathcal G}^{gen}_{14}({{\bf k},\omega})
          - \bar{\mathcal G}^{gen}_{34}({{\bf k},\omega})\right ] ~,\\
          \bar{\mathcal U}^{gen}_{21}({{\bf k},\omega}) =  -  2 \Re
\{\tilde{\Gamma}_{{{\rm cav}},\bf k}(\omega)\}
\frac{\kappa^{el~\star}_{{q}',{\bf
k}}}{\kappa^{ph~\star}_{{q},{\bf k}}}  \left [\bar{\mathcal
G}^{gen}_{21}({{\bf k},\omega})
          + \bar{\mathcal G}^{gen}_{41}({{\bf k},\omega})
          - \bar{\mathcal G}^{gen}_{23}({{\bf k},\omega})
          - \bar{\mathcal G}^{gen}_{43}({{\bf k},\omega})\right ] ~,\\
          \bar{\mathcal U}^{gen}_{22}({{\bf k},\omega}) =  1-  2 \Re
\{\tilde{\Gamma}_{{12},\bf k}(\omega)\} \left [\bar{\mathcal
G}^{gen}_{22}({{\bf k},\omega})
          + \bar{\mathcal G}^{gen}_{42}({{\bf k},\omega})
          - \bar{\mathcal G}^{gen}_{24}({{\bf k},\omega})
          - \bar{\mathcal G}^{gen}_{44}({{\bf k},\omega})\right ] ~.\\
\end{eqnarray}
%\twcolumngrid
 We have verified explicitly that the matrix
$\bar{\mathcal U}^{gen}({{\bf k},\omega})$ is unitary.
For the parameters used in this paper, $\bar{\mathcal U}^{gen}({{\bf k},\omega})
\simeq \bar{\mathcal U}({{\bf k},\omega})$, where $\bar{\mathcal U}({{\bf k},\omega})$
is the unitary matrix (\ref{U11}-\ref{U22}) in the absence of the bath
antiresonant terms.


\begin{thebibliography}{}
\bibitem{note_FFT} For the Fourier transforms, we have used the
definition $\tilde{A}(\omega) \equiv {\mathcal F}[A(t)](\omega)
\equiv ~\int_{-\infty}^{\infty} dt~ e^{i\omega t} A(t)~$. For the
Fourier transforms of hermitian conjugated operators, we will use
the simplified notation $ \tilde{A}^{\dagger}(-\omega) \equiv
{\mathcal F} [{A^{\dagger}(t)}](\omega) \equiv
(\tilde{A}(-\omega))^{\dagger} ~.$
\end{thebibliography}
\end{document}